\documentclass[lettersize,journal]{IEEEtran}
\usepackage{amsmath,amssymb,amsfonts}
\usepackage{graphicx}
\usepackage{textcomp}
\usepackage{cite}
\usepackage{picinpar}
\usepackage{url}
\usepackage[latin1,utf8]{inputenc}
\usepackage{colortbl}
\usepackage{soul}  
\usepackage{multirow}
\usepackage{pifont}
\usepackage{color}
\usepackage{alltt}
\usepackage{enumerate}
\usepackage{siunitx}
\usepackage{pbox}
\usepackage{makecell}
\usepackage{epstopdf}
\usepackage{bm}
\usepackage{tikz}
\usetikzlibrary{shapes.geometric, arrows.meta, positioning}
\usepackage{svg}
\usepackage{algorithm}
\usepackage{algpseudocode}
\usepackage{circuitikz}
\usepackage{mathtools}
\usepackage{pgfplots}
\pgfplotsset{compat=1.18}
\usepackage{placeins}
\usepackage{tikz}
\usetikzlibrary{positioning}
\usetikzlibrary{arrows.meta}

\usepackage{pdfpages}
\usepackage[hidelinks]{hyperref}

\begin{document}

\title{Generalized and Unified Power Flow Framework for Converter-Dominated AC/DC Power Grids}

\author{Hamin Chang, \IEEEmembership{Student Member, IEEE}, 
Fei Teng, \IEEEmembership{Senior Member, IEEE},\\
Gyu-Sub Lee, \IEEEmembership{Member, IEEE}, 
and Jae-Won Chang, \IEEEmembership{Member, IEEE}}

\markboth{SUBMITTED TO IEEE TRANSACTIONS ON POWER SYSTEMS}{Chang \MakeLowercase{\textit{et al.}}: Unified AC/DC Power Flow Framework}

\maketitle

\begin{abstract}
Modern power grids increasingly rely on power electronic interfaces, requiring steady-state power flow analysis to accommodate diverse operating modes and network structures.
Conventional approaches often rely on system-specific formulations and sequential approaches to handle such complexity.
However, these approaches may limit applicability to diverse grid configurations and lead to convergence and computational issues.
To address these limitations, this paper proposes a generalized and unified Newton--Raphson (NR) power flow framework for AC/DC hybrid grids.
For generalized operational modeling, conventional and droop operating modes of generators and interlinking converters are represented through a common set of coefficients.
For generalized structural modeling, models of AC transformers and DC/DC converters accommodate different transformer winding configurations and converter operating modes.
For a unified solution, augmented variables are introduced to incorporate droop deadbands, output limits, and converter station losses into a single NR formulation, eliminating the need for sequential outer loops.
To validate the effectiveness of the proposed framework, its results are compared with those obtained from existing power flow methods and electromagnetic transient simulations across transmission networks, distribution networks, and microgrids.
The implementation is made publicly available to facilitate its use in analyzing AC/DC hybrid grids.
\end{abstract}


\section{Introduction}
\IEEEPARstart{M}{odern} power grids are undergoing a major transition due to
the increasing penetration of distributed generators (DGs) and the rapid
growth of large-scale and concentrated loads, led by AI data centers
\cite{devries2023growing}. This transition is accompanied by the growing use of
power-electronic interfaces and increasing diversity in device operation
and network structure.
Operationally, generators and converter-interfaced resources
adopt various operating modes, including power, voltage,
frequency, and droop control~\cite{rathnayake2021grid}. Structurally, AC
networks are evolving from unidirectional to bidirectional
power flow, blurring the boundaries among transmission,
distribution, and microgrid systems \cite{jain2021integrated}. In parallel,
DC networks are expanding beyond conventional HVDC links
\cite{flourentzou2009vsc} into multi-terminal networks, distribution networks, and microgrids, becoming
integrated grids rather than auxiliary links.

These operational and structural changes have accelerated
the proliferation of AC/DC hybrid grids, where interlinking converters (ICs) serve
as controllable interfaces between AC and DC subgrids
\cite{unamuno2015hybrid}. ICs regulate power exchange through
multiple operating modes that enable coordinated operation of the two
subgrids \cite{gupta2017hybrid}. In addition, DC networks can encompass HVDC,
MVDC, and LVDC voltage levels, requiring DC/DC converters
between networks operating at different voltage levels \cite{paez2019overview}.
The power flow formulation for AC/DC hybrid grids must
represent the operating modes of
generators and ICs, along with the voltage transformation by
AC transformers and DC/DC converters. Beyond these operating modes,
generators and ICs require additional conditions, such as deadbands, that must be resolved
together with the network equations.

Existing AC/DC power flow methods developed to address these
requirements can be broadly classified into sequential and unified
approaches. Sequential
methods solve the AC and DC subgrid power flows separately and
iterate between them through an outer loop
\cite{beerten2012generalized,hamad2015sequential}. This structure is straightforward to implement, as it retains the
well-established AC power flow formulation and adds the DC network and
converter models through outer-loop procedures \cite{beerten2015matacdc,wang2014power}. However, sequential methods may suffer from convergence issues due to the coupling between the AC and DC subgrids through the ICs
\cite{zulkifli2026generalized}.
Beyond convergence issues, each subgrid power flow calculation uses AC/DC coupling variables
from the previous outer-loop iteration, which may be inconsistent with the subgrid's current state. Finally, accurately accounting for IC conversion
losses within the sequential AC/DC power flow formulation can be challenging
\cite{beerten2010sequential}.

To address the limitations of sequential approaches, unified methods
solve the AC and DC network equations in a single
Newton--Raphson (NR) formulation.
Early unified methods represent the converter as an
equivalent model, allowing the AC and DC subgrids to be solved within a
single power flow \cite{smed1991new,arboleya2012unified}. However, these
methods lack a generalized representation, both operationally and
structurally. Operationally, the conventional operating modes are not generalized,
nor is droop control considered. Structurally, these
methods are limited to specific network structures, such as railway
traction systems, and cannot be applied to general AC/DC hybrid grids.

To address these limitations, improved unified methods have extended the
generalization of grid component models, both operationally and structurally.
Operationally, the associated variables of generators and converters have
been classified in \cite{zulkifli2026generalized} to generalize the conventional
modes, which fix the voltage and phase angle, the active power and
voltage, or the active and reactive powers. Generator and IC droop control has also
been implemented for power sharing in
islanded AC/DC hybrid microgrids \cite{aprilia2019unified}, along with DC voltage
droop control in multi-terminal DC (MTDC) systems \cite{lei2017general}.
Nevertheless, this generalization is limited, as the conventional and droop
operating modes are represented separately.
Structurally,
voltage-transformation modeling has been generalized, with AC
transformers represented through controllable tap ratios and phase shifts
\cite{zulkifli2026generalized}, and different DC/DC converter configurations
included in hybrid AC/DC distribution network models \cite{javid2023unified}.
However, this generalization is also limited, as voltage-transformation
devices such as three-winding transformers and DC/DC converters with their
various operating modes are not considered within a single unified
formulation.

Beyond generalizing component models, a formulation must also incorporate elements whose effects depend on the current network
state, including droop deadbands, output limits, and
converter station losses. These elements are conventionally handled through
additional outer-loop procedures and are thus hereafter referred to as
outer-loop elements. Among the outer-loop elements, the converter station losses have been
incorporated into unified formulations
\cite{chai2016unified,lei2017general,ergun2019optimal}.
In contrast, the deadband and the output limits cause mode transitions that
depend on the network states, so conventional methods resolve these elements by
repeating the power flow through a sequential outer loop
\cite{wang2014power,chai2016unified}. As a result, these elements have not been
incorporated into the unified formulations, which are therefore not fully
unified.

As conventional methods are difficult to generalize and do not incorporate
the outer-loop elements into a unified formulation, this paper proposes a generalized and
unified power flow formulation with the following contributions:
\begin{enumerate}
    \item A generalized component representation is developed for AC/DC hybrid
    grids. For generalized operational modeling, the conventional and droop operating modes of
    generators and ICs are captured in a common coefficient-based structure.
    For generalized structural modeling, voltage transformation between multiple voltage levels is
    represented through models of AC transformers and DC/DC
    converters that accommodate different transformer winding configurations
    and converter operating modes.

    \item Augmented variables and corresponding constraints are introduced
to resolve the outer-loop elements together with the network equations,
requiring no sequential outer loop. These elements include the droop
deadband, the output limits, and the converter station
losses.

    \item All proposed models are integrated into a fully unified NR
    formulation capable of analyzing a wide range of converter-dominated AC/DC hybrid grids, from
    transmission and distribution networks to microgrids. The resulting
    implementation is made publicly available, enabling researchers to analyze
    diverse AC/DC hybrid grids \cite{chang2026unigrid}.
\end{enumerate}

\section{System Modeling}
\label{sec:gen_mode}

The proposed unified AC/DC power flow formulation includes general network and
load models, along with generalized component models for generators, ICs,
and voltage-transformation devices in the AC and DC subgrids. Let
$\mathcal{B}_{AC}$, $\mathcal{B}_{DC}$, and $\mathcal{B}_{IC}$ denote the sets
of AC buses, DC buses, and ICs, respectively. At each bus, $\mathcal{G}_i$, $\mathcal{L}_i$, and
$\mathcal{C}_i$ denote the sets of generators, loads, and ICs connected to AC
bus $i$, and $\mathcal{G}_k$, $\mathcal{L}_k$, and $\mathcal{C}_k$ those connected to DC bus $k$.
For DC/DC converters, $\mathcal{D}^{HV}_k$ and $\mathcal{D}^{LV}_k$ denote the sets of
converters whose high- and low-voltage bus is bus $k$, respectively.

\subsection{Network and Load Modeling}

The AC network is represented by a bus admittance matrix $\mathbf{Y}_{Bus}$, 
whose $(i,j)$th elements are defined as:
\begin{equation}
Y_{ij}(\omega) = \begin{cases}
{\displaystyle\sum_{\substack{j \in \mathcal{B}_{AC} \\ j\neq i}}} \left( y_{ij}(\omega) + \mathrm{j}\tfrac{B_{ij}(\omega)}{2} \right) + y_{sh,i}(\omega) & i = j \\[6pt]
-\,y_{ij}(\omega) & i \neq j
\end{cases}
\label{eq:ybus}
\end{equation}
where $y_{ij}(\omega)=(R_{ij}+\mathrm{j}X_{ij}(\omega))^{-1}$ and $B_{ij}(\omega)$ are the admittance and total line-charging susceptance of the branch between AC buses $i$ and $j$, and $y_{sh,i}(\omega)=g_{sh,i}+\mathrm{j}b_{sh,i}(\omega)$ is the shunt admittance at bus~$i$. Here, $\omega$ denotes the frequency of the AC subgrid containing bus~$i$, and $\mathbf{Y}_{Bus}$ depends on $\omega$ through the branch impedances and shunt susceptances.
The DC 
network is represented by a bus conductance matrix $\mathbf{G}_{Bus}$, 
whose $(k,l)$th elements are defined as:
\begin{equation}
G_{kl} = \begin{cases}
{\displaystyle\sum_{\substack{l \in \mathcal{B}_{DC} \\ l\neq k}}} g_{kl} & k = l \\[6pt]
-\,g_{kl} & k \neq l
\end{cases}
\label{eq:gbus}
\end{equation}
where $g_{kl} = R_{kl}^{-1}$ is the conductance of the branch between DC 
buses $k$ and $l$, with $R_{kl}$ denoting the branch resistance.

Both AC and DC loads are represented by the ZIP model, a combination of constant impedance (Z), constant current (I), and
constant power (P) components~\cite{aprilia2019unified}. For AC loads, the active and reactive
power demands of load $\lambda$ at bus $i$ depend on both voltage and
frequency and are expressed as:
\begin{equation}
\begin{split}
P_{LAC,\lambda} = P^0_{LAC,\lambda}&\left[Z_{p,\lambda}\left(\frac{|V_i|}{|V^0_i|}\right)^2 
+ I_{p,\lambda}\frac{|V_i|}{|V^0_i|} + P_{p,\lambda}\right]\\
&\times\left[1 + k_{pw,\lambda}(\omega - \omega^0)\right]
\end{split}
\label{eq:plac}
\end{equation}
\begin{equation}
\begin{split}
Q_{LAC,\lambda} = Q^0_{LAC,\lambda}&\left[Z_{q,\lambda}\left(\frac{|V_i|}{|V^0_i|}\right)^2 
+ I_{q,\lambda}\frac{|V_i|}{|V^0_i|} + Q_{q,\lambda}\right]\\
&\times\left[1 + k_{qw,\lambda}(\omega - \omega^0)\right],
\end{split}
\label{eq:qlac}
\end{equation}
where the superscript $0$ denotes the reference value of the corresponding variable, and $|V_i|$ denotes the voltage magnitude at bus $i$. The ZIP weighting coefficients for 
active power ($Z_{p,\lambda}$, $I_{p,\lambda}$, $P_{p,\lambda}$) and reactive power 
($Z_{q,\lambda}$, $I_{q,\lambda}$, $Q_{q,\lambda}$) each sum to unity. The coefficients $k_{pw,\lambda}$ and $k_{qw,\lambda}$ 
are the frequency sensitivities of the active and reactive load 
powers, respectively. For DC loads, the active-power demand of load $\lambda$ at bus $k$ depends on
voltage and is expressed as:

\begin{equation}
P_{LDC,\lambda} = P^0_{LDC,\lambda}\left[Z_{p,\lambda}\left(\frac{V_k}{V^0_k}\right)^2
+ I_{p,\lambda}\frac{V_k}{V^0_k} + P_{p,\lambda}\right],
\label{eq:pldc}
\end{equation}
where $V_k$ is the voltage at bus $k$, and the ZIP weighting coefficients
($Z_{p,\lambda}$, $I_{p,\lambda}$, $P_{p,\lambda}$) sum to unity.

\subsection{Generator Modeling}
\label{subsec:gen_mode}

\begin{figure}[!t]
\centering
\resizebox{\columnwidth}{!}{%
\begin{tikzpicture}[
    xscale=1.63,
    yscale=0.605,
    axis/.style={-{Stealth[length=3.2pt]}, line width=0.6pt},
    charline/.style={line width=0.9pt},
    modeline/.style={line width=0.7pt},
    rotarrow/.style={-{Stealth[length=3.6pt]}, line width=0.85pt, blue!65},
    every node/.style={font=\rmfamily\scriptsize}
]

\fill[green!8] (1.85,0) rectangle (2.55,4.32);
\draw[line width=0.35pt, green!50!black!45, dotted] (1.85,0) -- (1.85,4.32);
\draw[line width=0.35pt, green!50!black!45, dotted] (2.55,0) -- (2.55,4.32);
\draw[{Stealth[length=2.4pt]}-{Stealth[length=2.4pt]}, line width=0.45pt, green!50!black]
    (1.85,0.48) -- (2.55,0.48);
\node[font=\rmfamily\scriptsize, text=green!50!black, fill=green!8, inner sep=0.9pt]
    at (2.2,0.48) {$2V_{DB}$};

\draw[axis] (0,0) -- (4.75,0) node[below,xshift=-0.30cm] {$|V_i|$};
\draw[axis] (0,0) -- (0,5.0);
\node[left, font=\rmfamily\scriptsize] at (0,4.85) {$Q_{G,AC,g}$};

\draw[line width=0.4pt, dotted, red!70!black] (0,4.1) -- (4.3,4.1);
\draw[line width=0.4pt, dotted, red!70!black] (0,1.0) -- (4.3,1.0);

\draw[modeline, dashed] (2.2,1.0) -- (2.2,4.1);
\draw[line width=0.6pt] (2.2,0) -- (2.2,0.22);
\node[below, font=\rmfamily\scriptsize] at (2.2,0) {$|V^0_i|$};
\node[font=\rmfamily\scriptsize] at (2.2,4.5) {$m_{qAC,g}\!\approx\!\infty$};

\draw[modeline, dash dot] (0,2.55) -- (4.1,2.55);
\node[left, font=\rmfamily\scriptsize] at (0,2.55) {$Q^0_{G,AC,g}$};
\node[font=\rmfamily\scriptsize] at (3.72,2.90) {$m_{qAC,g}\!\approx\!0$};

\draw[charline, blue!70!black] (0.881,4.1) -- (1.85,2.55);
\draw[charline, green!50!black] (1.85,2.55) -- (2.55,2.55);
\draw[charline, blue!70!black] (2.55,2.55) -- (3.519,1.0);
\node[font=\rmfamily\scriptsize, text=blue!70!black,
      inner sep=0.8pt] at (1.482,3.689) {Droop};

\draw[rotarrow]
    (1.114,3.528) .. controls (0.814,3.178) and (0.854,2.828) .. (1.174,2.680);
\draw[rotarrow]
    (3.286,1.572) .. controls (3.586,1.922) and (3.546,2.272) .. (3.226,2.420);

\draw[modeline, dashed, red!75!black] (0.20,4.1) -- (2.2,4.1);
\draw[modeline, dashed, red!75!black] (2.2,1.0) -- (4.30,1.0);

\draw[modeline, dashed, red!75!black] (0.20,4.1) -- (2.2,4.1);
\draw[modeline, dashed, red!75!black] (2.2,1.0) -- (4.30,1.0);

\draw[charline, red!75!black] (0.20,4.1) -- (0.881,4.1);
\draw[charline, red!75!black] (3.519,1.0) -- (4.30,1.0);
\node[left, font=\rmfamily\scriptsize, text=red!75!black] at (0,4.1) {$Q^{\max}_{G,AC,g}$};
\node[left, font=\rmfamily\scriptsize, text=red!75!black] at (0,1.0) {$Q^{\min}_{G,AC,g}$};

\node[circle, fill=black, inner sep=0pt, minimum size=2.9pt] at (2.2,2.55) {};

\end{tikzpicture}
}
\caption{Generator operating-mode characteristics for the AC reactive-power
output, with the AC and DC active-power outputs taking the same form.}
\label{fig:droop_char}
\end{figure}
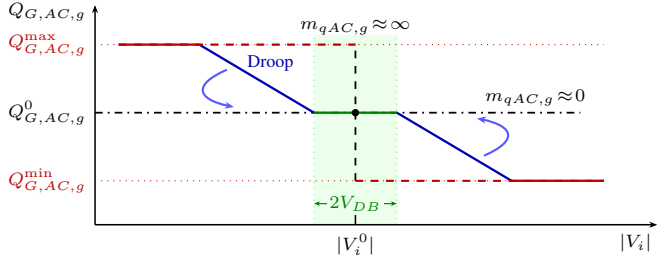

Generator modeling must represent the power outputs of AC- and DC-side 
generators under various operating modes. 
Beyond these operating modes, generator modeling must also represent the
elements whose effects depend on the power flow solution, which are the droop deadband and the output limits.
To address these requirements, a generalized coefficient-based formulation is 
proposed for both conventional and droop generator operating modes, with
the deadband and the output limits introduced as additional 
conditions. 
Fig.~\ref{fig:droop_char} shows the resulting operating-mode characteristics, 
using the AC reactive-power output as an example.

\subsubsection{Generalized Generator Operating Modes}
\label{subsec:gen_modes}

The power outputs of an AC-side generator $g$ 
at bus $i$ and a DC-side generator $g$ at bus $k$ are expressed as:
\begin{equation}
P_{G,AC,g} = P^0_{G,AC,g} + m_{pAC,g}(\omega^0 - \omega)
\label{eq:pgac}
\end{equation}
\begin{equation}
Q_{G,AC,g} = Q^0_{G,AC,g} + m_{qAC,g}(|V^0_i| - |V_i|)
\label{eq:qgac}
\end{equation}
\begin{equation}
P_{G,DC,g} = P^0_{G,DC,g} + m_{pDC,g}(V^0_k - V_k),
\label{eq:pgdc}
\end{equation}
where $m_{pAC,g}$, $m_{qAC,g}$, and $m_{pDC,g}$ denote the droop coefficients for the corresponding state variables.

Various generator operating modes can be implemented by
adjusting the droop coefficients in \eqref{eq:pgac}--\eqref{eq:pgdc}. 
As shown in Fig.~\ref{fig:droop_char}, setting any droop coefficient 
$m \approx \infty$ makes the slope infinite 
and fixes the state variable, whereas setting $m \approx 0$ makes the slope
zero and fixes the power output. 
Accordingly, the operating mode of an AC-side generator is determined by 
the coefficient pair $(m_{pAC,g}, m_{qAC,g})$ for its active- and reactive-power outputs,
and that of a DC-side generator by $m_{pDC,g}$ for its active-power output. 
Table~\ref{tab:generator_modes} summarizes the operating modes of AC- and 
DC-side generators. 
The constant frequency and voltage (slack), constant active power and 
voltage (PV), and constant active and reactive power (PQ) modes can be 
implemented for an AC-side generator. 
The constant DC voltage ($\mathrm{CV}_{DC}$) and constant power (CP) modes 
can be implemented for a DC-side generator. 
A finite intermediate value of $m$ instead makes the slope finite and 
implements the droop mode. 
Thus, a common set of coefficients represents both the conventional and droop modes within a single equation structure, rather than through separate formulations.

\subsubsection{Droop Deadband and Output Limits}
\label{subsec:gen_limits}

The generalized operating modes alone cannot represent the droop deadband
and the output limits. When the generator operates within the deadband or
reaches an output limit, the affected output becomes fixed and no longer varies
with the corresponding state variable, resulting in a mode transition.
The deadband and the output limits are therefore introduced as additional
conditions.
For the deadband, as shown in Fig.~\ref{fig:droop_char}, the generator output 
remains fixed at its set-point $Q^0_{G,AC,g}$ inside the deadband 
$|V^0_i| \pm V_{DB}$ and follows the droop characteristic outside the deadband. 
Here, $V_{DB}$ denotes the voltage deadband threshold. 
The output limits, in turn, switch the generator from the PV or droop mode to the PQ mode at either 
limit in Fig.~\ref{fig:droop_char}, with the output fixed at 
$Q^{\max}_{G,AC,g}$ or $Q^{\min}_{G,AC,g}$. 
These limits apply only to generators whose output powers are not already fixed by the operating mode, such as those in the PV and droop modes.
Similarly, the active-power outputs of AC- and DC-side generators in \eqref{eq:pgac} and \eqref{eq:pgdc}
follow the same characteristic shown in Fig.~\ref{fig:droop_char}, with 
$\omega_{DB}$ and $[P^{\min}_{G,AC,g}, P^{\max}_{G,AC,g}]$ for the AC-side 
generator, and $V_{DC,DB}$ and $[P^{\min}_{G,DC,g}, P^{\max}_{G,DC,g}]$ for 
the DC-side generator. 

\begin{table}[t]
\centering
\caption{Generator Operating Modes}
\label{tab:generator_modes}
\renewcommand{\arraystretch}{1.2}
\resizebox{\columnwidth}{!}{%
\begin{tabular}{lcc}
\hline
\multicolumn{3}{l}{\textit{AC-Side Generator Modes}} \\
\hline
Operating Mode & Coefficient Adjustment & Control Result \\
\hline
Slack & $m_{pAC,g} \approx\infty$, $m_{qAC,g} \approx\infty$ & Constant $\omega$, $|V_{i}|$ \\
PV & $m_{pAC,g} \approx 0$, $m_{qAC,g} \approx\infty$ & Constant $P_{G,AC,g}$, $|V_{i}|$ \\
PQ & $m_{pAC,g} \approx 0$, $m_{qAC,g} \approx 0$ & Constant $P_{G,AC,g}$, $Q_{G,AC,g}$ \\
Droop & Non-zero value for $m_{pAC,g}$, $m_{qAC,g}$ & $P$-$\omega$, $Q$-$|V_{i}|$ droop \\
\hline
\multicolumn{3}{l}{\textit{DC-Side Generator Modes}} \\
\hline
Operating Mode & Coefficient Adjustment & Control Result \\
\hline
CV$_{DC}$ & $m_{pDC,g} \approx\infty$ & Constant $V_{k}$ \\
CP & $m_{pDC,g} \approx 0$ & Constant $P_{G,DC,g}$ \\
Droop & Non-zero value for $m_{pDC,g}$ & $P$-$V_{k}$ droop \\
\hline
\end{tabular}%
}
\end{table}

However, whether the droop deadband or an output limit applies cannot be
determined before solving the power flow, because both depend on the network states. 
Both conditions must then be evaluated and fed back into the power flow, 
which changes the state variables and the generator outputs. 
Conventional methods therefore repeat this feedback through a sequential
outer loop, whereas the proposed formulation resolves both conditions
within the unified NR iteration, as described in Section~\ref{subsec:deadband}.

\subsection{Interlinking Converter Modeling}
\label{subsec:ic}

IC modeling must represent the power exchange between the AC and DC
subgrids under various operating modes. 
Beyond these operating modes, IC modeling must also represent the elements
whose effects depend on the power flow
solution, which are the converter station losses and the
output limits.
To address these requirements, a generalized coefficient-based formulation is
proposed for both conventional and droop IC operating modes, with the
losses and the output limits introduced as additional conditions. 

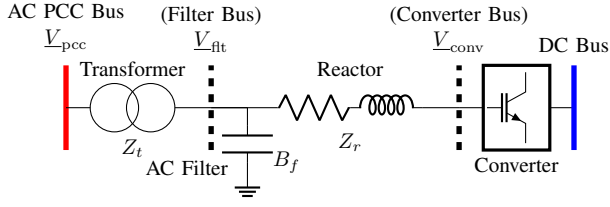
\begin{figure}[!t]
\centering
\begin{circuitikz}[american, scale=0.8, transform shape,
    font=\rmfamily]

\draw[line width=2pt, red] (0,-0.7) -- (0,0.7);
\node[above, font=\rmfamily, align=center] at (0,0.8) {AC PCC Bus\\$\underline{V}_{\mathrm{pcc}}$};

\draw (0.8,0) circle (0.4);
\draw (1.4,0) circle (0.4);
\draw (0,0) -- (0.4,0);
\draw (1.8,0) -- (2.4,0);
\node[above, font=\rmfamily] at (1.1,0.4) {Transformer};
\node[below, font=\rmfamily] at (1.1,-0.4) {$Z_t$};

\draw[line width=2pt, dashed] (2.4,-0.7) -- (2.4,0.7);
\node[above, font=\rmfamily, align=center] at (2.4,0.8) {(Filter Bus)\\$\underline{V}_{\mathrm{flt}}$};

\draw (2.4,0) -- (3.0,0);
\draw (3.0,0) -- (3.0,-0.3);
\draw (3.0,-0.3) to[C] (3.0,-0.9);
\draw (3.0,-0.9) node[ground]{};

\node[left, font=\rmfamily] at (2.8,-1) {AC Filter};
\node[right, font=\rmfamily] at (3.3,-0.9) {$B_f$};

\draw (2.4,0) -- (3.5,0);
\draw (3.5,0) to[R] (4.7,0);
\draw (4.7,0) to[L, cute] (5.9,0);
\draw (5.9,0) -- (6.5,0);
\node[above, font=\rmfamily] at (4.7,0.4) {Reactor};
\node[below, font=\rmfamily] at (4.7,-0.3) {$Z_r$};

\draw[line width=2pt, dashed] (6.5,-0.7) -- (6.5,0.7);
\node[above, font=\rmfamily, align=center] at (6.5,0.8) {(Converter Bus)\\$\underline{V}_{\mathrm{conv}}$};

\draw (6.5,0) -- (6.9,0);
\draw[line width=1.2pt] (6.9,-0.7) rectangle (8.0,0.7);
\draw (7.6,0) node[nigbt, scale=0.75, anchor=center]{};
\node[below, font=\rmfamily] at (7.45,-0.7) {Converter};
\draw (8.0,0) -- (8.4,0);

\draw[line width=2pt, blue] (8.4,-0.7) -- (8.4,0.7);
\node[above, font=\rmfamily] at (8.4,0.8) {DC Bus};

\end{circuitikz}
\caption{Internal structure of the IC between the PCC and DC buses.}
\label{fig:vsc_structure}
\end{figure}

Fig.~\ref{fig:vsc_structure} shows each IC modeled as a converter station
that connects the AC-side point of common coupling (PCC) bus to the DC
bus. Each station includes an internal
circuit on its AC side, with a transformer, filter, and reactor. To represent this circuit, the filter and converter buses are introduced as
virtual AC buses, and the bus voltages associated with each IC $c \in \mathcal{B}_{IC}$
are denoted using subscripts that indicate their roles, such as $V_{\mathrm{pcc},c}$ for the
PCC bus voltage.

\subsubsection{Generalized IC Operating Modes}
The PCC-side active-power
exchange is expressed as:
\begin{equation}
P_{IC,AC,c} = -m_{\omega,c}\hat{\omega} 
+ m_{V_{DC},c}\hat{V}_{DC,c} + P^0_{IC,AC,c},
\label{eq:pic_ac_droop}
\end{equation}
and the reactive-power exchange as:
\begin{equation}
Q_{IC,AC,c} = m_{qIC,c}
\left(|V^0_{\mathrm{pcc},c}| 
- |V_{\mathrm{pcc},c}|\right) + Q^0_{IC,AC,c},
\label{eq:qic_ac}
\end{equation}
where $m_{\omega,c}$, $m_{V_{DC},c}$, and $m_{qIC,c}$ denote the droop coefficients for the corresponding variables.
Note that the power is defined as positive when flowing toward the AC side. 
Here, $\hat{\omega}$ and $\hat{V}_{DC,c}$ denote 
the normalized AC frequency and DC voltage at the DC bus of converter 
$c$, respectively, defined as:
\begin{equation}
\hat{\omega} = 
\frac{\omega - 0.5(\omega_{\max}+\omega_{\min})}
{0.5(\omega_{\max}-\omega_{\min})}
\label{eq:omega_hat}
\end{equation}
\begin{equation}
\hat{V}_{DC,c} = 
\frac{V_{DC,c} 
- 0.5(V_{DC,\max,c}+V_{DC,\min,c})}
{0.5(V_{DC,\max,c}-V_{DC,\min,c})},
\label{eq:vdc_norm}
\end{equation}
where $\omega_{\max}$ and $\omega_{\min}$ are the upper and lower bounds of
the AC frequency, and $V_{DC,\max,c}$ and $V_{DC,\min,c}$ those of
the DC bus voltage. Generally, the bounds are set to satisfy
$\omega^0 = 0.5(\omega_{\max}+\omega_{\min})$ and
$V^0_{DC,c} = 0.5(V_{DC,\max,c}+V_{DC,\min,c})$, so that both variables
span the same range from $-1$ to $+1$ and become zero at the reference (nominal) values.

The opposite signs of the two droop terms in \eqref{eq:pic_ac_droop} ensure
that each supports its corresponding subgrid. Specifically, a frequency
drop increases the power injected into the AC subgrid, and a DC voltage
drop increases that into the DC subgrid. In AC/DC microgrids, both droop coefficients $m_{\omega,c}$ and $m_{V_{DC},c}$ are often nonzero. This allows the IC to provide mutual support between the AC and DC subgrids by adjusting the power exchange in response to deviations in both the normalized frequency and the DC voltage. In contrast, in transmission and distribution systems, one of these coefficients is typically set to zero so that the IC supports either the AC or the DC subgrid. 
In either case, the operating mode of
the IC is determined by the coefficient pair $(m_{\omega,c}, m_{V_{DC},c})$ for active power and 
by $m_{qIC,c}$ for reactive power. 
Table~\ref{tab:ic_modes} summarizes the operating modes of the IC, which fall into three categories: (i) the grid-controlling modes---the constant
frequency (CF) and constant AC voltage ($\mathrm{CV}_{AC}$) modes for the AC subgrid
and the $\mathrm{CV}_{DC}$ mode for the DC subgrid; (ii) the power-controlling modes---the constant active power (CP)
and constant reactive power (CQ) modes; and (iii) the droop mode---the adjustment of the
power exchange in response to deviations in the frequency, AC voltage, or DC
voltage.
Thus, a common set of coefficients represents both the conventional 
and droop modes of the IC within a single equation structure.

\begin{table}[!t]
\centering
\caption{Interlinking Converter Operating Modes}
\label{tab:ic_modes}
\renewcommand{\arraystretch}{1.2}
\resizebox{\columnwidth}{!}{%
\begin{tabular}{lcc}
\hline
Operating Mode & Coefficient Adjustment & Control Result \\
\hline
\multicolumn{3}{l}{\textit{Active Power Control}} \\
CV$_{DC}$ & $m_{\omega,c} \approx 0$, $m_{V_{DC},c} \approx\infty$ & Constant $V_{DC,c}$ \\
CF & $m_{\omega,c} \approx\infty$, $m_{V_{DC},c} \approx 0$ & Constant $\omega$ \\
CP & $m_{\omega,c} \approx 0$, $m_{V_{DC},c} \approx 0$ & Constant $P_{IC,AC,c}$ \\
Droop & Non-zero value for $m_{\omega,c}$ and/or $m_{V_{DC},c}$ & $P$-$\omega$/$V_{DC,c}$ droop \\
\hline
\multicolumn{3}{l}{\textit{Reactive Power Control}} \\
CV$_{AC}$ & $m_{qIC,c} \approx\infty$ & Constant $|V_{\mathrm{pcc},c}|$ \\
CQ & $m_{qIC,c} \approx 0$ & Constant $Q_{IC,AC,c}$ \\
Droop & Non-zero value for $m_{qIC,c}$ & $Q$-$|V_{\mathrm{pcc},c}|$ droop \\
\hline
\end{tabular}%
}
\end{table}

\subsubsection{Converter Station Losses}

Each IC is connected to the grid at the PCC bus through the internal circuit.
As shown in Fig.~\ref{fig:vsc_structure}, \eqref{eq:pic_ac_droop} and \eqref{eq:qic_ac} define the PCC-side powers of IC $c$, so the internal losses must be added to them.
These losses occur in the internal circuit, which is represented by
(i) the transformer admittance
$y_{t,c}=Z_{t,c}^{-1}$ between the PCC and filter buses, (ii) the reactor
admittance $y_{r,c}=Z_{r,c}^{-1}$ between the filter and converter buses, and
(iii) the filter susceptance $B_{f,c}$ shunt-connected at the filter bus.
These elements are incorporated into
$\mathbf{Y}_{Bus}$, and their losses are also computed explicitly to relate the
PCC-side powers to the converter-bus injections.

To obtain the internal losses, the complex power consumed in the internal circuit
is computed from the internal-circuit bus voltages $\underline{V}_{\mathrm{r},c}=|V_{\mathrm{r},c}|e^{\mathrm{j}\delta_{\mathrm{r},c}}$
in Fig.~\ref{fig:vsc_structure} as:
\begin{equation}
\begin{split}
S_{int,c} ={}&
y_{t,c}^{*}
\left|
\underline{V}_{\mathrm{pcc},c}
-
\underline{V}_{\mathrm{flt},c}
\right|^2 \\
&+
y_{r,c}^{*}
\left|
\underline{V}_{\mathrm{flt},c}
-
\underline{V}_{\mathrm{conv},c}
\right|^2
-
\mathrm{j}B_{f,c}
\left|
\underline{V}_{\mathrm{flt},c}
\right|^2 ,
\end{split}
\label{eq:sint}
\end{equation}
where $\delta_{\mathrm{r},c}$ is the phase angle, $\mathrm{r} \in \{\mathrm{pcc},\mathrm{flt},\mathrm{conv}\}$ denotes
the PCC, filter, and converter buses of IC $c$, and $P_{int,loss,c} = \operatorname{Re}\{S_{int,c}\}$ and
$Q_{int,loss,c} = \operatorname{Im}\{S_{int,c}\}$ are the active- and
reactive-power losses.
Thus, the actual converter-bus injections of IC $c$ are defined as:
\begin{equation}
\tilde{P}_{IC,AC,c} = P_{IC,AC,c} + P_{int,loss,c}
\label{eq:u_pvsc}
\end{equation}
\begin{equation}
\tilde{Q}_{IC,AC,c} = Q_{IC,AC,c} + Q_{int,loss,c}.
\label{eq:u_qvsc}
\end{equation}

Beyond the internal-circuit losses in \eqref{eq:u_pvsc} and
\eqref{eq:u_qvsc}, the IC's own conversion loss is
modeled as a quadratic function of the
converter current magnitude~\cite{beerten2012generalized}:
\begin{equation}
P_{IC,loss,c} = a_c + b_c|I_{IC,c}| + c_c|I_{IC,c}|^2,
\label{eq:vsc_loss}
\end{equation}
where $a_c$, $b_c$, and $c_c$ are the no-load, linear, and quadratic loss
coefficients of IC $c$, respectively. Here, the converter current magnitude
follows from the converter-bus injections as:
\begin{equation}
|I_{IC,c}| =
\frac{
\sqrt{\tilde{P}_{IC,AC,c}^2 + \tilde{Q}_{IC,AC,c}^2}
}{
|V_{\mathrm{conv},c}|
}.
\label{eq:ivsc}
\end{equation}
The conversion loss is then included in the IC's DC-side injection
$P_{IC,DC,c}$, obtained from $\tilde{P}_{IC,AC,c}$ as:
\begin{equation}
P_{IC,DC,c}=-\tilde{P}_{IC,AC,c}-P_{IC,loss,c}.
\label{eq:pic_dc}
\end{equation}

Thus, the converter station losses cannot be determined before solving the
power flow, because the internal losses depend on the internal-circuit bus voltages 
and the conversion loss on the converter current in \eqref{eq:ivsc}. 
The conversion loss must then be evaluated from \eqref{eq:ivsc} and 
\eqref{eq:vsc_loss} and fed back into the power flow, which changes
the current in \eqref{eq:ivsc}.
Conventional methods therefore repeat this feedback through a sequential
outer loop, whereas the proposed formulation resolves these
losses within the NR iteration, as described in Section~\ref{subsec:vsc}.

\subsubsection{Output Limits}

The converter current rating $I^{\max}_{IC,c}$ and the nominal power rating
$S^{N}_{IC,c}$ of IC $c$ define its apparent-power capability $S^{\max}_{IC,c}$ as:
\begin{equation}
S^{\max}_{IC,c} = \min\!\left(S^{N}_{IC,c},\; 
I^{\max}_{IC,c}\,|V_{\mathrm{conv},c}|\right).
\label{eq:smax_vsc}
\end{equation}
Active power is prioritized over reactive power for determining the output limits in accordance with common practice~\cite{beerten2012generalized,beerten2015matacdc}. The active-power limit ($P^{\max}_{IC,c}$) is therefore $P^{\max}_{IC,c}=S^{\max}_{IC,c}$, and the reactive-power limit
$Q^{\max}_{IC,c}$ is expressed as:
\begin{equation}
Q^{\max}_{IC,c} = \sqrt{\left(S^{\max}_{IC,c}\right)^{2}-\tilde{P}_{IC,AC,c}^{2}}.
\label{eq:qmax_vsc}
\end{equation}
When the active-power injection reaches $P^{\max}_{IC,c}$, the IC no longer follows the operating mode in~\eqref{eq:pic_ac_droop} and fixes the injection at $\pm P^{\max}_{IC,c}$, resulting in $Q^{\max}_{IC,c}=0$ in~\eqref{eq:qmax_vsc}. For reactive power, when the injection reaches $Q^{\max}_{IC,c}$, the IC no longer follows the operating mode in~\eqref{eq:qic_ac} and fixes the injection at $\pm Q^{\max}_{IC,c}$. The operation in which the active- or reactive-power injection is fixed at its limit is referred to as capability-bound operation.

The operating mode of the IC cannot be determined before solving the power flow.
This is because the transition to capability-bound operation depends on the converter-bus voltage 
magnitude in \eqref{eq:smax_vsc} and the active-power injection in \eqref{eq:qmax_vsc}. 
The output limits must then be evaluated and fed back into the power flow, 
which changes the converter-bus voltage magnitude and the active-power injection. 
Conventional methods therefore repeat this feedback through a sequential
outer loop, whereas the proposed formulation resolves the output limits
within the NR iteration, as described in Section~\ref{subsec:vsc}.

\subsection{Voltage-Transformation Device Modeling}
\label{subsec:vtd}

Voltage-transformation device modeling covers the AC transformers and DC/DC converters that interconnect multiple voltage levels.
Both two- and three-winding AC transformers are embedded into the AC bus admittance matrix, and a generalized coefficient-based formulation is proposed for the DC/DC converter operating modes.

\subsubsection{AC Transformers}
A two-winding transformer is modeled as a branch from bus $f$ to bus $t$, 
with an ideal transformer of tap ratio $\tau$ and phase shift angle 
$\theta_{shift}$ connected in series at bus $f$~\cite{zimmerman2011matpower}.
The tap ratio and phase shift enter $\Delta Y_{ij}$, which modifies the 
$(i,j)$th element of $\mathbf{Y}_{Bus}$ in \eqref{eq:ybus}, for 
$i,j \in \{f,t\}$:
\begin{equation}
\begin{bmatrix}
\Delta Y_{ff} & \Delta Y_{ft} \\
\Delta Y_{tf} & \Delta Y_{tt}
\end{bmatrix}
=
\begin{bmatrix}
\left(y_s + \mathrm{j}\frac{B_{ft}}{2}\right)\frac{1}{\tau^2} &
-y_s \frac{1}{\tau e^{-\mathrm{j}\theta_{shift}}} \\
-y_s \frac{1}{\tau e^{\mathrm{j}\theta_{shift}}} &
y_s + \mathrm{j}\frac{B_{ft}}{2}
\end{bmatrix}
\label{eq:yft}
\end{equation}
where $y_s = (R_{ft} + \mathrm{j}X_{ft})^{-1}$ is the admittance of the
transformer branch, and $B_{ft}$ is its total shunt susceptance. 
For a three-winding transformer interconnecting the HV, MV, and LV voltage levels, the windings are modeled as three two-winding transformers connected at an auxiliary bus \cite{thurner2018pandapower}. Each winding
enters $\mathbf{Y}_{Bus}$ as a branch following the same form as \eqref{eq:yft}, and
the auxiliary bus is included in $\mathbf{Y}_{Bus}$ as an AC bus. The winding admittances and the additional unknowns of the auxiliary bus are detailed in Appendix A.

\subsubsection{DC/DC Converters}
Each DC/DC converter $d$ connects a high- and a low-voltage DC bus, 
with terminal voltages $V_{\mathrm{HV},d}$ and $V_{\mathrm{LV},d}$, respectively. 
Similar to the IC active-power formulation in 
\eqref{eq:pic_ac_droop}, the active power injected into the low-voltage 
DC bus of converter $d$ is expressed as:
\begin{equation}
P_{conv,LV,d}
=
m_{HV,d}\hat{V}_{\mathrm{HV},d}
-
m_{LV,d}\hat{V}_{\mathrm{LV},d}
+
P^0_{conv,LV,d},
\label{eq:pconv_lv}
\end{equation}
where $m_{HV,d}$ and $m_{LV,d}$ denote the droop coefficients for the 
corresponding variables, and $\hat{V}_{\mathrm{HV},d}$ and $\hat{V}_{\mathrm{LV},d}$ are the terminal
voltages $V_{\mathrm{HV},d}$ and $V_{\mathrm{LV},d}$ normalized as in \eqref{eq:vdc_norm}. 
On the high-voltage side, the active-power injection is defined by 
the power balance as:
\begin{equation}
P_{conv,HV,d}
=
-P_{conv,LV,d}.
\label{eq:pconv_hv}
\end{equation}
The DC/DC converters are modeled as lossless in the present formulation.
Conversion losses and output limits, when required, can be incorporated
in the same manner as the IC conversion loss in \eqref{eq:vsc_loss} and
the IC output limits.

The opposite signs of the $m_{HV,d}$ and $m_{LV,d}$ droop terms in \eqref{eq:pconv_lv} ensure 
that these terms provide voltage support to the high- and low-voltage DC buses, respectively. Specifically, a voltage drop on 
either bus increases the power injected into that bus. 
In DC microgrids \cite{qiu2026distributed}, both droop coefficients $m_{HV,d}$ and 
$m_{LV,d}$ are often nonzero, allowing the DC/DC converter to respond to 
voltage deviations on both sides. In contrast, in DC transmission and 
distribution systems \cite{ye2023power}, one of these coefficients is 
typically set to zero so that the converter supports either the high- or 
the low-voltage side. In either case, the operating mode of the DC/DC converter is 
determined by the coefficient pair $(m_{HV,d}, m_{LV,d})$. 
Table~\ref{tab:dcdc_modes} summarizes the operating modes of the DC/DC
converter, which fall into three categories: (i) the grid-controlling modes---the constant-voltage modes ($\mathrm{CV}_{HV}$ and $\mathrm{CV}_{LV}$) for the high- and low-voltage DC buses; (ii) the power-controlling mode---the constant power (CP) mode; and (iii) the droop mode---the adjustment of the power transfer in response to deviations in the high- or low-voltage DC bus voltage.
Thus, a common set of coefficients represents both the conventional and
droop modes of the DC/DC converter within a single equation structure.

\begin{table}[!t]
\centering
\caption{DC/DC Converter Operating Modes}
\label{tab:dcdc_modes}
\renewcommand{\arraystretch}{1.2}
\resizebox{\columnwidth}{!}{%
\begin{tabular}{lcc}
\hline
Operating Mode & Coefficient Adjustment & Control Result \\
\hline
CV$_{HV}$ & $m_{HV,d} \approx\infty$, $m_{LV,d} \approx 0$ & Constant $V_{\mathrm{HV},d}$ \\
CV$_{LV}$ & $m_{HV,d} \approx 0$, $m_{LV,d} \approx\infty$ & Constant $V_{\mathrm{LV},d}$ \\
CP & $m_{HV,d} \approx 0$, $m_{LV,d} \approx 0$ & Constant $P_{conv,LV,d}$ \\
Droop & Non-zero value for $m_{HV,d}$ and/or $m_{LV,d}$ & $P$-$V_{\mathrm{HV},d}$/$V_{\mathrm{LV},d}$ droop \\
\hline
\end{tabular}%
}
\end{table}

\section{Unified Newton--Raphson Formulation}
\label{sec:nr}

The models of Section~\ref{sec:gen_mode} cover the AC and DC networks and loads, with generalized component models for generators, ICs, and voltage-transformation devices.
These models are then integrated into a unified AC/DC power flow, which is solved by a baseline NR formulation. However, the droop
deadband, the output limits, and the converter station losses still require an additional outer loop outside the
NR iteration.
This section therefore extends the baseline into a fully unified
augmented NR formulation requiring no outer loop.

\subsection{Baseline NR Formulation}
\label{subsec:baseline}

The baseline formulation applies the NR method to the coupled AC/DC network
equations. The network state variables are gathered in the baseline state vector
$\mathbf{x}_{base}$, as given by:
\begin{equation}
\mathbf{x}_{base} =
[|\mathbf{V}_{AC}|,\, \boldsymbol{\omega},\, \boldsymbol{\delta},\, \mathbf{V}_{DC}]^T,
\label{eq:state}
\end{equation}
where $|\mathbf{V}_{AC}|$, $\boldsymbol{\omega}$, $\boldsymbol{\delta}$, and
$\mathbf{V}_{DC}$ denote the AC voltage magnitudes, AC subgrid frequencies,
phase angles, and DC bus voltages, respectively. Note that, in the presence of asynchronous AC subgrids, each AC subgrid has its own frequency, fixed at $\omega^0$ if it contains a slack bus or an IC in CF mode and otherwise solved as a state variable.
In addition, one bus angle in each AC subgrid is fixed as the reference. For AC buses $i$
and $j$, the calculated injections are
obtained from $\mathbf{Y}_{Bus}$ in \eqref{eq:ybus} as:
\begin{equation}
\begin{aligned}
P_{cal,AC,i} &= |V_i| \sum_{j\in\mathcal{B}_{AC}} |Y_{ij}(\omega)||V_j|
\cos(\delta_i-\delta_j-\theta_{ij}) \\
Q_{cal,AC,i} &= |V_i| \sum_{j\in\mathcal{B}_{AC}} |Y_{ij}(\omega)||V_j|
\sin(\delta_i-\delta_j-\theta_{ij})
\end{aligned}
\label{eq:pcal_ac}
\end{equation}
where $\theta_{ij}=\angle Y_{ij}(\omega)$. For DC
buses $k$ and $l$, the calculated injection is obtained from
$\mathbf{G}_{Bus}$ in \eqref{eq:gbus} as:
\begin{equation}
P_{cal,DC,k} = V_k \sum_{l\in\mathcal{B}_{DC}} G_{kl}V_l.
\label{eq:pcal_dc}
\end{equation}

The specified injections at each bus are formed by generators, loads,
ICs, and DC/DC converters, as given by:
\begin{equation}
P_{spec,AC,i} = \sum_{g\in\mathcal{G}_i} P_{G,AC,g}
+ \sum_{c\in\mathcal{C}_i} P_{IC,AC,c}
- \sum_{\lambda\in\mathcal{L}_i} P_{LAC,\lambda}
\label{eq:psch_ac}
\end{equation}
\begin{equation}
Q_{spec,AC,i} = \sum_{g\in\mathcal{G}_i} Q_{G,AC,g}
+ \sum_{c\in\mathcal{C}_i} Q_{IC,AC,c}
- \sum_{\lambda\in\mathcal{L}_i} Q_{LAC,\lambda}
\label{eq:qsch_ac}
\end{equation}
\begin{equation}
\begin{aligned}
P_{spec,DC,k} ={}& \sum_{g\in\mathcal{G}_k} P_{G,DC,g}
+ \sum_{c\in\mathcal{C}_k}P_{IC,DC,c}
- \sum_{\lambda\in\mathcal{L}_k}P_{LDC,\lambda} \\
&+ \sum_{d\in\mathcal{D}^{LV}_k}P_{conv,LV,d}
+ \sum_{d\in\mathcal{D}^{HV}_k}P_{conv,HV,d}.
\end{aligned}
\label{eq:psch_dc}
\end{equation}
The generator terms follow \eqref{eq:pgac}--\eqref{eq:pgdc}, the load terms
\eqref{eq:plac}--\eqref{eq:pldc}, the IC terms
\eqref{eq:pic_ac_droop}--\eqref{eq:qic_ac}, and the DC/DC converter terms
\eqref{eq:pconv_lv}--\eqref{eq:pconv_hv}. The baseline formulation does not consider the converter station losses in~\eqref{eq:pic_dc}, so the IC's DC-side injection is modeled as
$P_{IC,DC,c}=-P_{IC,AC,c}$ until these losses are incorporated in
Section~\ref{subsec:vsc}.

The calculated injections \eqref{eq:pcal_ac}--\eqref{eq:pcal_dc} are
determined by the network state through $\mathbf{Y}_{Bus}$ and
$\mathbf{G}_{Bus}$, whereas the specified injections
\eqref{eq:psch_ac}--\eqref{eq:psch_dc} are defined by the component models.
The power flow equations require the two injections to be equal at each bus.
Accordingly, the baseline mismatch vector is defined as:
\begin{equation}
\mathbf{F}_{base} =
\begin{bmatrix}
\mathbf{P}_{spec,AC}-\mathbf{P}_{cal,AC} \\
\mathbf{Q}_{spec,AC}-\mathbf{Q}_{cal,AC} \\
\mathbf{P}_{spec,DC}-\mathbf{P}_{cal,DC}
\end{bmatrix}.
\label{eq:mismatch}
\end{equation}

The nonlinear equations $\mathbf{F}_{base}(\mathbf{x}_{base})=\mathbf{0}$ are
solved iteratively by the NR method. Each iteration requires the
baseline power-flow Jacobian
$\mathbf{J}_{base}=\partial\mathbf{F}_{base}/\partial\mathbf{x}_{base}$.
At iteration $\nu$, the state vector is updated as:
\begin{equation}
\mathbf{x}_{base}^{(\nu+1)} = \mathbf{x}_{base}^{(\nu)} -
\left(\mathbf{J}_{base}^{(\nu)}\right)^{-1}\mathbf{F}_{base}^{(\nu)},
\label{eq:nr_update}
\end{equation}
until the maximum mismatch falls below a tolerance $\varepsilon$.

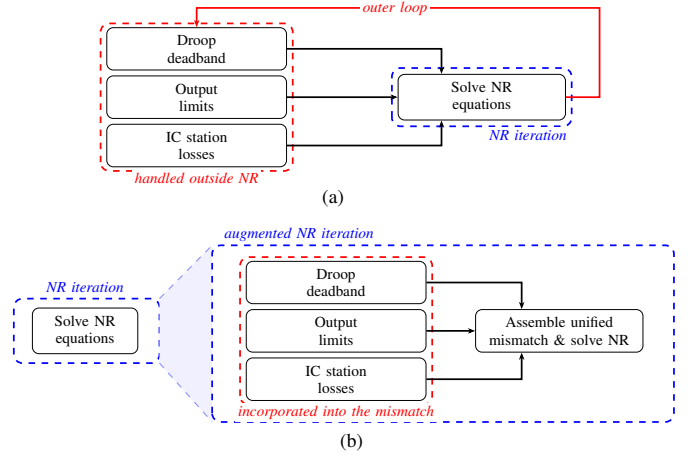
\begin{figure}[!t]
\centering
\resizebox{\columnwidth}{!}{%
\begin{tikzpicture}[
    box/.style={rectangle, rounded corners=3pt, draw, minimum width=2.9cm,
        minimum height=0.62cm, font=\scriptsize, align=center},
    widebox/.style={rectangle, rounded corners=3pt, draw, minimum width=2.7cm,
        minimum height=0.62cm, font=\scriptsize, align=center},
    arr/.style={-{Stealth[length=3.4pt]}, thick},
    loop/.style={-{Stealth[length=3.4pt]}, thick},
    group/.style={dashed, rounded corners=4pt, line width=0.8pt},
    glabel/.style={font=\scriptsize\itshape, fill=white, inner sep=1.2pt},
    plabel/.style={font=\small}
]
\node[box] (ae1) at (2.5,8.72) {Droop\\deadband};
\node[box] (ae2) at (2.5,7.94) {Output\\limits};
\node[box] (ae3) at (2.5,7.16) {IC station\\losses};
\node[widebox] (anr) at (7.1,7.94) {Solve NR\\equations};
\draw[group, red] (0.95,6.75) rectangle (4.05,9.13);
\node[glabel, text=red] at (2.5,6.63) {handled outside NR};
\draw[group, blue] (5.65,7.47) rectangle (8.55,8.41);
\node[glabel, text=blue, anchor=east] at (8.52,7.33) {NR iteration};
\draw[arr] (ae1.east) -| (anr.150);
\draw[arr] (ae2.east) -- (anr.west);
\draw[arr] (ae3.east) -| (anr.210);
\draw[loop, red] (anr.east) -- (9.0,7.94) -- (9.0,9.38) -- (2.5,9.38) -- (ae1.north);
\node[glabel, text=red] at (5.7,9.38) {outer loop};
\node[plabel] at (4.7,6.31) {(a)};
\node[rectangle, rounded corners=3pt, draw, minimum width=1.7cm, minimum height=0.62cm,
      font=\scriptsize, align=center] (bnr) at (0.7,4.17) {Solve NR\\equations};
\draw[group, blue] (-0.45,3.65) rectangle (1.9,4.69);
\node[glabel, text=blue] at (0.7,4.87) {NR iteration};
\fill[blue!6] (1.9,4.69) -- (2.75,5.55) -- (2.75,2.70) -- (1.9,3.65) -- cycle;
\draw[blue!45, dashed, line width=0.5pt] (1.9,4.69) -- (2.75,5.55);
\draw[blue!45, dashed, line width=0.5pt] (1.9,3.65) -- (2.75,2.70);
\node[box] (be1) at (4.75,4.95) {Droop\\deadband};
\node[box] (be2) at (4.75,4.17) {Output\\limits};
\node[box] (be3) at (4.75,3.39) {IC station\\losses};
\node[widebox] (bsolve) at (8.35,4.17) {Assemble unified\\mismatch \& solve NR};
\draw[group, blue] (2.75,2.70) rectangle (10.2,5.55);
\draw[group, red] (3.20,2.98) rectangle (6.30,5.36);
\draw[arr] (be1.east) -| (bsolve.150);
\draw[arr] (be2.east) -- (bsolve.west);
\draw[arr] (be3.east) -| (bsolve.210);
\node[glabel, text=blue, anchor=south west] at (2.9,5.57) {augmented NR iteration};
\node[glabel, text=red] at (4.75,2.86) {incorporated into the mismatch};
\node[plabel] at (5.0,2.37) {(b)};
\end{tikzpicture}
}
\vspace{-1.3em}
\caption{Outer-loop elements: (a) conventionally handled outside the NR iteration and (b) integrated into the proposed augmented NR formulation.}
\label{fig:db_flow}
\end{figure}

\subsection{Integration of Outer-Loop Elements}
\label{subsec:outer}

The baseline formulation does not represent the outer-loop
elements, which are the droop deadband, the output
limits, and the converter station losses. These elements depend on the network
states and are therefore difficult to incorporate directly into the baseline NR
formulation. The proposed formulation introduces augmented variables and
corresponding constraints to resolve these elements together with the network
equations. Fig.~\ref{fig:db_flow}(a) shows how the outer-loop elements are conventionally
handled outside the NR iteration, whereas Fig.~\ref{fig:db_flow}(b) shows their
integration into the proposed augmented NR formulation. 

\subsubsection{Concept of the Augmented NR Formulation}
\label{subsec:concept}

As shown in Fig.~\ref{fig:db_flow}(a), the outer-loop elements are
conventionally handled through a sequential outer loop. The associated
quantities are updated outside the NR iteration and then fed back into the
network equations. To avoid this outer loop, the proposed
formulation carries these quantities as additional unknowns, denoted by the
augmented variables $\mathbf{x}_{aug}$. Each augmented variable is determined by
the characteristic of its corresponding device (i.e., a generator or an IC) as a
function of the network states and other augmented variables, in the general
form $\mathbf{x}_{aug}=\mathbf{T}(\mathbf{x}_{aug},\mathbf{x}_{base})$. This
relationship is incorporated into the NR iteration through the constraint function
$\mathbf{F}_{aug}(\mathbf{x}_{aug},\mathbf{x}_{base})=\mathbf{x}_{aug}-\mathbf{T}(\mathbf{x}_{aug},\mathbf{x}_{base})$.
Thus, setting $\mathbf{F}_{aug}=\mathbf{0}$ ensures that every augmented
variable satisfies the characteristic of its corresponding device. The baseline state vector $\mathbf{x}_{base}$ is then
appended with $\mathbf{x}_{aug}$, and the constraint vector $\mathbf{F}_{aug}$ is also appended to the
baseline mismatch vector $\mathbf{F}_{base}$. Therefore, as shown in
Fig.~\ref{fig:db_flow}(b), the network equations and augmented constraints are
solved simultaneously within the NR iteration, without a sequential outer loop.

\subsubsection{Droop Deadband and Output Limits}
\label{subsec:deadband}

As modeled in Section~\ref{subsec:gen_limits}, the droop deadband and the output
limits modify the characteristic of \eqref{eq:pgac}--\eqref{eq:pgdc}, as shown in Fig.~\ref{fig:droop_char}. Thus, the generator
outputs in \eqref{eq:psch_ac}--\eqref{eq:psch_dc} are replaced by
augmented variables ($\tilde{P}_{G,AC,g}$ for $P_{G,AC,g}$, $\tilde{Q}_{G,AC,g}$ for
$Q_{G,AC,g}$, and $\tilde{P}_{G,DC,g}$ for $P_{G,DC,g}$), collected as:
\begin{equation}
\mathbf{x}_{G} =
[\tilde{\mathbf{P}}_{G,AC},\, \tilde{\mathbf{Q}}_{G,AC},\, \tilde{\mathbf{P}}_{G,DC}]^T .
\label{eq:u_db}
\end{equation}
The constraint functions for each generator represent the
characteristics modified by the deadband and limit conditions. For the AC
reactive-power output, the constraint function is defined as:
\begin{equation}
\begin{aligned}
&F_{Q,AC,g} = \\
&\begin{cases}
\tilde{Q}_{G,AC,g} - \big(Q^0_{G,AC,g} + m_{qAC,g}(|V^0_i| - V_{DB} - |V_i|)\big), \\
\hspace{3.3cm}(|V_i| < |V^0_i| - V_{DB}) \\
\tilde{Q}_{G,AC,g} - \big(Q^0_{G,AC,g} + m_{qAC,g}(|V^0_i| + V_{DB} - |V_i|)\big), \\
\hspace{3.3cm}(|V_i| > |V^0_i| + V_{DB}) \\
\tilde{Q}_{G,AC,g} - Q^0_{G,AC,g}, \quad (\big||V^0_i| - |V_i|\big| \leq V_{DB}) \\
\tilde{Q}_{G,AC,g} - Q^{\max}_{G,AC,g}, \quad \text{(upper limit)} \\
\tilde{Q}_{G,AC,g} - Q^{\min}_{G,AC,g}, \quad \text{(lower limit)}
\end{cases}
\end{aligned}
\label{eq:fu_qac}
\end{equation}
Here, the output-limit conditions take precedence over the deadband and droop
conditions.

Similarly, the AC and DC active-power constraints $\mathbf{F}_{P,AC}$ and
$\mathbf{F}_{P,DC}$ follow the same form as \eqref{eq:fu_qac}, with the
corresponding deadband and output limits. The three constraint groups
$\mathbf{F}_{P,AC}$, $\mathbf{F}_{Q,AC}$, and $\mathbf{F}_{P,DC}$ are then
collected as:
\begin{equation}
\mathbf{F}_{G} =
[\mathbf{F}_{P,AC},\, \mathbf{F}_{Q,AC},\, \mathbf{F}_{P,DC}]^T .
\label{eq:f_db}
\end{equation}
The baseline state vector $\mathbf{x}_{base}$ is then appended with $\mathbf{x}_{G}$, and the
generator constraint vector $\mathbf{F}_{G}$ is also appended to the baseline mismatch vector $\mathbf{F}_{base}$.
Accordingly, the augmented Jacobian includes the blocks
$\partial\mathbf{F}_{base}/\partial\mathbf{x}_{G}$,
$\partial\mathbf{F}_{G}/\partial\mathbf{x}_{base}$, and
$\partial\mathbf{F}_{G}/\partial\mathbf{x}_{G}$. The resulting augmented system
is solved in Section~\ref{subsec:augmented}.
If the IC droop characteristics in \eqref{eq:pic_ac_droop} and
\eqref{eq:qic_ac} include a deadband, the deadband can be incorporated as
specified in \eqref{eq:fu_qac}.

\subsubsection{Converter Station Losses and Output Limits}
\label{subsec:vsc}
As modeled in Section~\ref{subsec:ic}, each IC's actual converter-bus
injections differ from its PCC-side powers in
\eqref{eq:pic_ac_droop}--\eqref{eq:qic_ac} by the active- and reactive-power
losses of the internal circuit. These injections cannot be determined before solving the power flow, because the internal losses depend on the internal-circuit bus voltages. To resolve this, these injections
in \eqref{eq:psch_ac}--\eqref{eq:qsch_ac} are replaced by augmented
variables ($\tilde{P}_{IC,AC,c}$ for $P_{IC,AC,c}$ and
$\tilde{Q}_{IC,AC,c}$ for $Q_{IC,AC,c}$), collected as:
\begin{equation}
\mathbf{x}_{IC}=[\tilde{\mathbf{P}}_{IC},\,\tilde{\mathbf{Q}}_{IC}]^T.
\label{eq:u_vsc}
\end{equation}

The constraint functions for each IC account for the
internal-circuit losses in \eqref{eq:u_pvsc}--\eqref{eq:u_qvsc} and the output limits in \eqref{eq:smax_vsc}--\eqref{eq:qmax_vsc}, and are defined as:
\begin{equation}
\begin{aligned}
&F_{P,IC,c} = \\
&\begin{cases}
\tilde{P}_{IC,AC,c}-(P_{IC,AC,c}+P_{int,loss,c}), \\
\hspace{1.2cm}(|P_{IC,AC,c}+P_{int,loss,c}| \leq P^{\max}_{IC,c}) \\
\tilde{P}_{IC,AC,c}-\operatorname{sign}(P_{IC,AC,c}+P_{int,loss,c})\,P^{\max}_{IC,c}, \\
\hspace{4.4cm}\text{(otherwise)}
\end{cases}
\end{aligned}
\label{eq:f_pvsc}
\end{equation}
\begin{equation}
\begin{aligned}
&F_{Q,IC,c} = \\
&\begin{cases}
\tilde{Q}_{IC,AC,c}-(Q_{IC,AC,c}+Q_{int,loss,c}), \\
\hspace{1.2cm}(|Q_{IC,AC,c}+Q_{int,loss,c}| \leq Q^{\max}_{IC,c}) \\
\tilde{Q}_{IC,AC,c}-\operatorname{sign}(Q_{IC,AC,c}+Q_{int,loss,c})\,Q^{\max}_{IC,c}, \\
\hspace{4.4cm}\text{(otherwise)}
\end{cases}
\end{aligned}
\label{eq:f_qvsc}
\end{equation}
The two constraint groups $\mathbf{F}_{P,IC}$ and
$\mathbf{F}_{Q,IC}$ are then collected as:
\begin{equation}
\mathbf{F}_{IC}=[\mathbf{F}_{P,IC},\,\mathbf{F}_{Q,IC}]^T.
\label{eq:f_vsc}
\end{equation}
The baseline state vector $\mathbf{x}_{base}$ is then appended with $\mathbf{x}_{IC}$, and the
IC constraint vector $\mathbf{F}_{IC}$ is also appended to the baseline mismatch vector $\mathbf{F}_{base}$.
Accordingly, the augmented Jacobian includes the
blocks $\partial\mathbf{F}_{base}/\partial\mathbf{x}_{IC}$,
$\partial\mathbf{F}_{IC}/\partial\mathbf{x}_{base}$, and
$\partial\mathbf{F}_{IC}/\partial\mathbf{x}_{IC}$.

\subsection{Augmented NR Formulation}
\label{subsec:augmented}

To solve the outer-loop elements together with the network equations, the
augmented variables and constraints of Section~\ref{subsec:outer} are combined
into a single NR system.
The baseline state vector $\mathbf{x}_{base}$ and the augmented variables $\mathbf{x}_{G}$ and
$\mathbf{x}_{IC}$ are gathered in the unified state vector
$\mathbf{x}$, as given by:
\begin{equation}
\mathbf{x}=[\mathbf{x}_{base},\,\mathbf{x}_{G},\,\mathbf{x}_{IC}]^T.
\label{eq:state_aug}
\end{equation}
The baseline mismatch vector $\mathbf{F}_{base}$ and the constraint vectors $\mathbf{F}_{G}$ and
$\mathbf{F}_{IC}$ are gathered in the unified mismatch vector
$\mathbf{F}$, as given by:
\begin{equation}
\mathbf{F}=[\mathbf{F}_{base},\,\mathbf{F}_{G},\,\mathbf{F}_{IC}]^T.
\label{eq:mismatch_aug}
\end{equation}
The augmented system is solved by NR
iteration with the augmented Jacobian expressed as:
\begin{equation}
\mathbf{J}=
\begin{bmatrix}
\dfrac{\partial \mathbf{F}_{base}}{\partial \mathbf{x}_{base}} &
\dfrac{\partial \mathbf{F}_{base}}{\partial \mathbf{x}_{G}} &
\dfrac{\partial \mathbf{F}_{base}}{\partial \mathbf{x}_{IC}} \\[10pt]
\dfrac{\partial \mathbf{F}_{G}}{\partial \mathbf{x}_{base}} &
\dfrac{\partial \mathbf{F}_{G}}{\partial \mathbf{x}_{G}} &
\dfrac{\partial \mathbf{F}_{G}}{\partial \mathbf{x}_{IC}} \\[10pt]
\dfrac{\partial \mathbf{F}_{IC}}{\partial \mathbf{x}_{base}} &
\dfrac{\partial \mathbf{F}_{IC}}{\partial \mathbf{x}_{G}} &
\dfrac{\partial \mathbf{F}_{IC}}{\partial \mathbf{x}_{IC}}
\end{bmatrix}.
\label{eq:jacobian}
\end{equation}
Note that the rows and columns of the unified mismatch vector
\eqref{eq:mismatch_aug} and the augmented Jacobian \eqref{eq:jacobian} are modified according to
the operating modes and mode transitions of generators, ICs, and DC/DC
converters. An operating mode that
fixes a state variable removes the column and the corresponding augmented
constraint or mismatch row, whereas both are restored when a generator or IC reaches an output limit.
Furthermore, the blocks $\partial\mathbf{F}_{base}/\partial\mathbf{x}_{base}$ and
$\partial\mathbf{F}_{base}/\partial\mathbf{x}_{IC}$ in~\eqref{eq:jacobian} contain
the derivatives of the conversion loss $P_{IC,loss,c}$, which are particularly challenging to derive.
The resulting rows and columns and the derivation of every block, together with the chain rule for the partial derivatives of the IC conversion loss, are provided in
Appendix B.

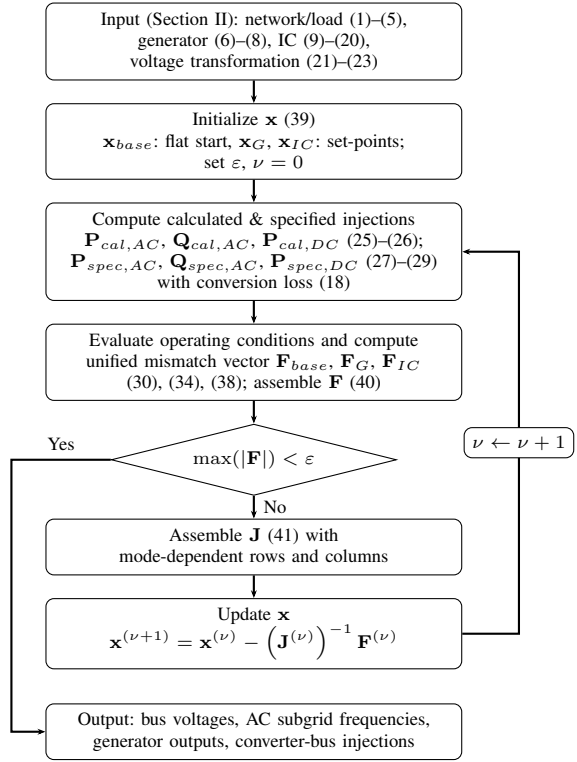
\begin{figure}[!t]
\centering
\begin{tikzpicture}[
    node distance=0.3cm,
    box/.style={rectangle, rounded corners=3pt, draw, minimum width=5.5cm,
        minimum height=0.6cm, font=\scriptsize, align=center},
    decision/.style={diamond, draw, font=\scriptsize, align=center, aspect=4},
    arr/.style={-{Stealth[length=3pt]}, thick}
]

\node[box] (input) {Input (Section~\ref{sec:gen_mode}): network/load \eqref{eq:ybus}--\eqref{eq:pldc},\\
    generator \eqref{eq:pgac}--\eqref{eq:pgdc}, IC \eqref{eq:pic_ac_droop}--\eqref{eq:qmax_vsc},\\
    voltage transformation \eqref{eq:yft}--\eqref{eq:pconv_hv}};

\node[box, below=of input] (init) {Initialize $\mathbf{x}$ \eqref{eq:state_aug}\\
    $\mathbf{x}_{base}$: flat start, $\mathbf{x}_{G}$, $\mathbf{x}_{IC}$: set-points;\\
    set $\varepsilon$, $\nu=0$};

\node[box, below=of init] (inj) {Compute calculated \& specified injections\\
    $\mathbf{P}_{cal,AC}$, $\mathbf{Q}_{cal,AC}$, $\mathbf{P}_{cal,DC}$ \eqref{eq:pcal_ac}--\eqref{eq:pcal_dc};\\
    $\mathbf{P}_{spec,AC}$, $\mathbf{Q}_{spec,AC}$, $\mathbf{P}_{spec,DC}$ \eqref{eq:psch_ac}--\eqref{eq:psch_dc}\\
    with conversion loss \eqref{eq:pic_dc}};

\node[box, below=of inj] (mismatch) {Evaluate operating conditions and compute\\
    unified mismatch vector $\mathbf{F}_{base}$, $\mathbf{F}_{G}$, $\mathbf{F}_{IC}$\\
    \eqref{eq:mismatch}, \eqref{eq:f_db}, \eqref{eq:f_vsc}; assemble $\mathbf{F}$ \eqref{eq:mismatch_aug}};

\node[decision, below=0.3cm of mismatch] (converge)
    {$\max(|\mathbf{F}|)<\varepsilon$};

\node[box, below=0.3cm of converge] (jacobian) {Assemble $\mathbf{J}$ \eqref{eq:jacobian} with\\
    mode-dependent rows and columns};

\node[box, below=of jacobian] (update) {Update $\mathbf{x}$\\
    $\mathbf{x}^{(\nu+1)} =
    \mathbf{x}^{(\nu)} -
    \left(\mathbf{J}^{(\nu)}\right)^{-1}
    \mathbf{F}^{(\nu)}$};

\node[box, below=0.45cm of update] (output) {Output: bus voltages, AC subgrid frequencies,\\
    generator outputs, converter-bus injections};

\draw[arr] (input) -- (init);
\draw[arr] (init) -- (inj);
\draw[arr] (inj) -- (mismatch);
\draw[arr] (mismatch) -- (converge);
\draw[arr] (converge) -- node[right, font=\scriptsize]{No} (jacobian);
\draw[arr] (jacobian) -- (update);
\draw[arr] (converge.west) -- node[above, font=\scriptsize, pos=0.5]{Yes}
    ($(input.west |- converge.west)+(-0.45,0)$) |- (output.west);
\draw[arr] (update.east) -- ++(0.75,0) coordinate(right)
    |- node[draw, thin, rounded corners=3pt, fill=white, font=\scriptsize, inner sep=3pt, pos=0.25]{$\nu \leftarrow \nu+1$} (inj.east);

\end{tikzpicture}
\caption{Overall flowchart of the proposed augmented NR formulation.}
\label{fig:flowchart}
\end{figure}

Fig.~\ref{fig:flowchart} shows the overall procedure of the proposed
augmented NR formulation, which uses the models of
Section~\ref{sec:gen_mode} as its input. The unified state vector
$\mathbf{x}$ is initialized with $\mathbf{x}_{base}$ at a flat start,
whereas $\mathbf{x}_{G}$ and $\mathbf{x}_{IC}$ begin at their set-points. The remaining steps follow the
baseline NR procedure of Section~\ref{subsec:baseline}, applied to the
unified state vector and the unified mismatch vector. Thus, the network equations
and the outer-loop elements are solved simultaneously within this fully unified
NR formulation.

\section{Case Studies}
\label{sec:case_study}
The proposed generalized and unified formulation is applied to various test systems that span a wide range of scales, from transmission and distribution systems to islanded microgrids.

\subsection{Modified IEEE RTS-96 AC/DC Transmission System}
\label{subsec:case24}

For validation on a transmission system, the proposed formulation is applied
to a modified three-zone IEEE RTS-96
system~\cite{grigg1999rts,beerten2012generalized}.
The solutions of the proposed formulation are compared with those of the
sequential AC/DC power flow method MatACDC~\cite{beerten2015matacdc} to
validate the formulation and to show its advantages relative to the sequential
method.
Fig.~\ref{fig:case24_acdc_system} shows the modified RTS-96 system, in which
a three-terminal and a four-terminal meshed MTDC grid interlink three
asynchronous AC subgrids through seven ICs. Each AC subgrid contains one slack bus that fixes its frequency at $\omega^0$ (60\,Hz), and the other AC-side generators operate in the PV mode. Two scenarios (A1 and A2) are examined
to test the generalized IC operating-mode formulation. Each IC's operating
mode is listed in Fig.~\ref{fig:case24_acdc_system}.
The two scenarios differ in the active-power control of the ICs. Scenario A1
applies constant DC-voltage control ($\mathrm{CV}_{DC}$), in which IC1 and IC4
fix the voltage of their respective DC grids while the remaining five ICs
hold constant power (CP).
Scenario A2 applies $P$--$V_{DC}$ droop control to all seven ICs.
Since MatACDC does not support reactive-power droop control, IC2 regulates its PCC voltage ($\mathrm{CV}_{AC}$) in both scenarios,
while the remaining six ICs hold constant reactive power (CQ). The comparison
shows the voltage magnitudes at every AC and DC bus, together with the
AC-side voltage angles and the IC power injections.
Fig.~\ref{fig:case24_validation} shows the deviations from MatACDC in both
scenarios. The solutions of the two methods are almost identical, with maximum deviations of
$8.0 \times 10^{-6}$\,p.u. in magnitude, $3.1 \times 10^{-3}\,\mathrm{deg}$ in angle, and
$5.7 \times 10^{-4}$\,p.u. in IC power.

\begin{figure}[!t]
\centering
\resizebox{\columnwidth}{!}{%
\begin{tikzpicture}[
    font=\rmfamily,
    aczone1/.style={draw=red!70!black, thick, rounded corners=4pt,
        fill=red!5, minimum width=1.75cm, minimum height=2.05cm, align=center},
    aczone2/.style={draw=red!70!black, thick, rounded corners=4pt,
        fill=red!5, minimum width=1.75cm, minimum height=2.05cm, align=center},
    aczone3/.style={draw=red!70!black, thick, rounded corners=4pt,
        fill=red!5, minimum width=1.95cm, minimum height=0.85cm, align=center},
    dcbusblue/.style={rectangle, draw=blue, fill=blue,
        minimum size=4.6pt, inner sep=0pt},
    dcbusgreen/.style={rectangle, draw=green!55!black, fill=green!55!black,
        minimum size=4.6pt, inner sep=0pt},
    dclineblue/.style={line width=0.65pt, blue},
    dclinegreen/.style={line width=0.65pt, green!55!black},
    acline/.style={line width=0.45pt, black!70},
    rtsbus/.style={circle, draw=red!70!black, fill=red!70!black,
        opacity=0.36, minimum size=2.0pt, inner sep=0pt},
    rtsline/.style={red!70!black, line width=0.34pt, opacity=0.34},
    rtsgen/.style={circle, draw=red!70!black, fill=white,
        line width=0.34pt, opacity=0.72, minimum size=5.0pt,
        inner sep=0pt, font=\tiny},
    boxlabel/.style={font=\bfseries\scriptsize},
    buslabel/.style={font=\tiny},
    vscname/.style={font=\scriptsize, align=center, inner sep=0.8pt,
        fill=white, fill opacity=0.90, text opacity=1},
    vscmode/.style={font=\scriptsize, align=center, inner sep=0.8pt,
        fill=white, fill opacity=0.90, text opacity=1}
]

\newcommand{\vscicon}[3]{%
    \node[draw=black, thick, fill=gray!8, minimum width=0.36cm,
        minimum height=0.36cm, inner sep=0pt] (#1) at (#2,#3) {};
    \draw[line width=0.35pt] (#2-0.11,#3-0.10) -- (#2+0.10,#3+0.10);
    \draw[line width=0.35pt] (#2-0.10,#3+0.07) -- (#2-0.03,#3+0.07);
    \draw[line width=0.35pt] (#2+0.03,#3-0.07) -- (#2+0.10,#3-0.07);
}

\newcommand{\rtswatermark}[3]{%
    \begin{scope}[shift={(#1,#2)}, scale=#3]
        \draw[rtsline] (-0.12,0.54) -- (0.34,0.47)
            -- (0.56,0.18) -- (0.18,0.10) -- (-0.34,0.13) -- (-0.12,0.54);
        \draw[rtsline] (-0.34,0.13) -- (-0.53,-0.18) -- (-0.18,-0.36)
            -- (0.30,-0.30) -- (0.58,-0.52);
        \draw[rtsline] (-0.12,0.54) -- (0.18,0.10) -- (0.30,-0.30);
        \draw[rtsline] (0.34,0.47) -- (0.18,0.10) -- (-0.18,-0.36);
        \draw[rtsline] (-0.53,-0.18) -- (-0.54,-0.54) -- (-0.18,-0.36);
        \draw[rtsline] (0.56,0.18) -- (0.30,-0.30);

        \node[rtsgen] at (-0.58,0.52) {G};
        \draw[rtsline] (-0.58,0.52) -- (-0.34,0.13);

        \foreach \x/\y in {
            -0.12/0.54, 0.34/0.47, 0.56/0.18,
            0.18/0.10, -0.34/0.13, -0.53/-0.18, -0.18/-0.36,
            0.30/-0.30, 0.58/-0.52, -0.54/-0.54}
            \node[rtsbus] at (\x,\y) {};
    \end{scope}
}

\node[aczone1] (z1) at (-4.95,0.25) {};
\node[font=\bfseries\footnotesize, text=red!70!black] at (-4.95,1.02) {AC Zone 1};
\node[font=\scriptsize] at (-4.95,0.72) {RTS--96};
\node[font=\tiny] at (-4.95,0.49) {Buses 101--124};
\rtswatermark{-4.95}{-0.18}{0.78}

\node[aczone2] (z2) at (5.15,0.10) {};
\node[font=\bfseries\footnotesize, text=red!70!black] at (5.15,0.87) {AC Zone 2};
\node[font=\scriptsize] at (5.15,0.57) {RTS--96};
\node[font=\tiny] at (5.15,0.34) {Buses 201--224};
\rtswatermark{5.15}{-0.33}{0.78}

\node[aczone3] (z3) at (-0.55,-2.75) {
    {\footnotesize\bfseries\textcolor{red!70!black}{AC Zone 3}}\\[-1pt]
    \scriptsize Offshore / Wind\\[-1pt]
    \tiny Buses 301--302
};

\coordinate (z2botVSC2) at (4.70,-0.925);
\coordinate (z2botVSC6) at (5.15,-0.925);

\node[font=\bfseries\small, text=blue, align=center] at (-2.05,1.82)
    {3-terminal\\HVDC-MTDC (150\,kV)};

\node[font=\bfseries\small, text=green!55!black, align=center] at (2.08,1.82)
    {4-terminal meshed\\HVDC-MTDC (300\,kV)};

\node[dcbusblue] (dc1) at (-2.70,0.90) {};
\node[dcbusblue] (dc2) at (-2.05,-0.05) {};
\node[dcbusblue] (dc3) at (-1.40,0.90) {};

\node[buslabel] at (-2.70,1.12) {DC1};
\node[buslabel] at (-1.78,-0.15) {DC2};
\node[buslabel] at (-1.40,1.12) {DC3};

\draw[dclineblue] (dc1) -- (dc3);
\draw[dclineblue] (dc2) -- (dc3);

\node[dcbusgreen] (dc4) at (1.40,0.90) {};
\node[dcbusgreen] (dc5) at (1.40,-0.55) {};
\node[dcbusgreen] (dc6) at (2.75,0.90) {};
\node[dcbusgreen] (dc7) at (2.75,-0.55) {};

\node[buslabel] at (1.40,1.12) {DC4};
\node[buslabel] at (1.74,-0.74) {DC5};
\node[buslabel] at (2.75,1.12) {DC6};
\node[buslabel] at (2.75,-0.78) {DC7};

\draw[dclinegreen] (dc4) -- (dc5);
\draw[dclinegreen] (dc4) -- (dc6);
\draw[dclinegreen] (dc4) -- (dc7);
\draw[dclinegreen] (dc5) -- (dc7);
\draw[dclinegreen] (dc6) -- (dc7);

\vscicon{vsc1}{-3.28}{0.90}
\node[vscname] at (-3.28,1.24) {IC1};

\vscicon{vsc2}{-2.05}{-1.25}
\node[vscname] at (-1.58,-1.23) {IC2};

\vscicon{vsc3}{-0.82}{0.90}
\node[vscname] at (-0.82,1.24) {IC3};

\vscicon{vsc4}{0.82}{0.90}
\node[vscname] at (0.82,1.24) {IC4};

\vscicon{vsc5}{0.82}{-0.55}
\node[vscname] at (0.82,-0.18) {IC5};

\vscicon{vsc6}{3.45}{0.90}
\node[vscname] at (3.45,1.24) {IC6};

\vscicon{vsc7}{3.45}{-0.55}
\node[vscname] at (3.45,-0.20) {IC7};

\draw[dclineblue] (vsc1) -- (dc1);
\draw[dclineblue] (dc2) -- (vsc2);
\draw[dclineblue] (dc3) -- (vsc3);
\draw[dclinegreen] (vsc4) -- (dc4);
\draw[dclinegreen] (dc5) -- (vsc5);

\draw[dclinegreen] (dc6) -- (vsc6);
\draw[dclinegreen] (dc7) -- (vsc7);

\coordinate (ac107) at (-3.85,0.90);
\coordinate (ac113) at (-3.85,-0.55);
\coordinate (ac204) at (-2.05,-1.75);
\coordinate (ac301) at (-0.55,-1.65);
\coordinate (ac123) at (4.10,0.90);
\coordinate (ac215) at (4.10,-1.35);
\coordinate (ac217) at (4.10,-0.55);

\draw[acline] (z1.east |- ac107) -- (ac107) -- (vsc1);

\draw[acline] (z1.east |- ac113) -- (ac113) -- (-0.25,-0.55) -- (-0.25,0.90) -- (vsc4);

\draw[acline] (vsc3) -- (-0.55,0.90) -- (ac301) -- (z3.north);

\draw[acline] (vsc2) -- (ac204) -- (z2botVSC2 |- ac204) -- (z2botVSC2);

\draw[acline] (vsc6) -- (ac123) -- (z2.west |- ac123);

\draw[acline] (vsc5) -- (0.18,-0.55) -- (0.18,-1.0) -- (-4.95,-1.0) -- (z1.south);

\draw[acline] (vsc7) -- (ac217) -- (z2.west |- ac217);

\node[anchor=north, font=\rmfamily\scriptsize, inner sep=1pt]
      at ([yshift=-3mm]current bounding box.south) {%
  \renewcommand{\arraystretch}{1.15}%
  \begin{tabular}{@{}l@{\ \ }*{6}{c@{\ \ }}c@{}}
    \hline
    & IC1 & IC2 & IC3 & IC4 & IC5 & IC6 & IC7 \\
    \hline
    A1 & CV\textsubscript{DC}/CQ & CP/CV\textsubscript{AC} & CP/CQ & CV\textsubscript{DC}/CQ & CP/CQ & CP/CQ & CP/CQ \\
    A2 & Droop/CQ & Droop/CV\textsubscript{AC} & Droop/CQ & Droop/CQ & Droop/CQ & Droop/CQ & Droop/CQ \\
    \hline
  \end{tabular}%
};
\end{tikzpicture}%
}
\caption{Modified three-zone IEEE RTS-96 transmission system.}
\label{fig:case24_acdc_system}
\end{figure}
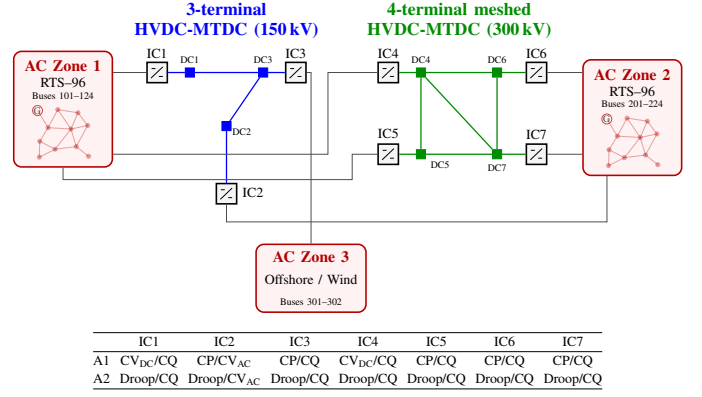
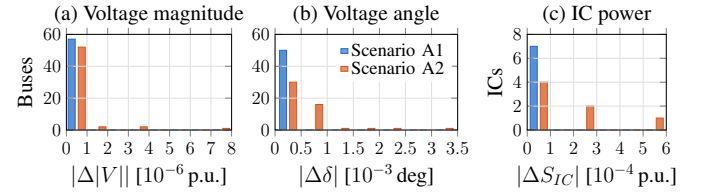
\begin{figure}[!t]
\centering
\definecolor{s1c}{rgb}{0.12,0.42,0.80}
\definecolor{s2c}{rgb}{0.85,0.33,0.10}
\resizebox{\columnwidth}{!}{%
\begin{tikzpicture}
\pgfplotsset{every axis/.append style={
  width=3.9cm, height=2.2cm, scale only axis,
  ybar, bar width=4.6pt, bar shift auto, ymin=0, ymax=60, ytick={0,20,40,60},
  tick label style={font=\large}, label style={font=\Large},
  title style={font=\Large, yshift=-3pt},
  ymajorgrids, xmajorgrids, grid style={gray!25}, axis on top,
  legend style={font=\large, draw=none, fill=white, fill opacity=0.85,
                text opacity=1, at={(0.97,0.97)}, anchor=north east, row sep=-2pt},
  legend image code/.code={\draw[#1] (0cm,-0.055cm) rectangle (0.16cm,0.055cm);}}}

\begin{axis}[name=a, width=3.8cm, xmin=0, xmax=8, xtick={0,1,2,3,4,5,6,7,8},
  xlabel={$|\Delta|V||$ [$10^{-6}$\,p.u.]}, ylabel={Buses}, title={(a) Voltage magnitude}]
\addplot[fill=s1c!75, draw=s1c] coordinates {(0.5,57) (1.5,0) (2.5,0) (3.5,0) (4.5,0) (5.5,0) (6.5,0) (7.5,0)};
\addplot[fill=s2c!75, draw=s2c] coordinates {(0.5,52) (1.5,2) (2.5,0) (3.5,2) (4.5,0) (5.5,0) (6.5,0) (7.5,1)};
\end{axis}

\begin{axis}[name=b, at={(a.east)}, anchor=west, xshift=10mm, width=4.2cm,
  xmin=0, xmax=3.5, xtick={0,0.5,1,1.5,2,2.5,3,3.5},
  xlabel={$|\Delta\delta|$ [$10^{-3}$\,deg]}, title={(b) Voltage angle}]
\addplot[fill=s1c!75, draw=s1c] coordinates {(0.25,50) (0.75,0) (1.25,0) (1.75,0) (2.25,0) (2.75,0) (3.25,0)};
\addplot[fill=s2c!75, draw=s2c] coordinates {(0.25,30) (0.75,16) (1.25,1) (1.75,1) (2.25,1) (2.75,0) (3.25,1)};
\legend{Scenario A1, Scenario A2}
\end{axis}

\begin{axis}[name=c, at={(b.east)}, anchor=west, xshift=16mm, width=3.2cm,
  xmin=0, xmax=6, xtick={0,1,2,3,4,5,6}, ymax=8, ytick={0,2,4,6,8},
  xlabel={$|\Delta S_{IC}|$ [$10^{-4}$\,p.u.]}, ylabel={ICs}, title={(c) IC power}]
\addplot[fill=s1c!75, draw=s1c] coordinates {(0.5,7) (1.5,0) (2.5,0) (3.5,0) (4.5,0) (5.5,0)};
\addplot[fill=s2c!75, draw=s2c] coordinates {(0.5,4) (1.5,0) (2.5,2) (3.5,0) (4.5,0) (5.5,1)};
\end{axis}
\end{tikzpicture}}
\caption{Histograms of the deviations between the proposed and MatACDC solutions:
(a) voltage magnitude, (b) voltage angle, and (c) IC power.}
\label{fig:case24_validation}
\end{figure}

\begin{table}[!t]
\centering
\caption{Convergence and Computational Time as the Power Exchange through All Seven ICs
Is Scaled}
\label{tab:case24_scaling}
\footnotesize
\setlength{\tabcolsep}{0pt}
\renewcommand{\arraystretch}{1.2}
\begin{tabular*}{\columnwidth}{@{}l@{\extracolsep{\fill}}cc@{\hspace{8pt}}ccc@{}}
\hline\hline
 & \multicolumn{2}{c}{\textbf{Proposed} (unified)} & \multicolumn{3}{c}{\textbf{MatACDC} (sequential)} \\
\cline{2-3}\cline{4-6}
Scaling & Iterations & Time [ms] & Outer & AC NR & Time [ms] \\
factor & & & iterations & iterations & \\
\hline
$\times$1   & 4 & 61.5 & 4 & 23 & 474.7 \\
$\times$13  & 6 & 63.1 & 4 & 32 & 475.4 \\
$\times$14  & 7 & 64.8 & \multicolumn{3}{c}{no convergence} \\
\hline\hline
\end{tabular*}
\end{table}

To verify its superior convergence characteristics and computational
efficiency, the proposed formulation is compared with the sequential method
under strong AC/DC coupling.
Table~\ref{tab:case24_scaling}
summarizes the convergence and computational time as the active-power
set-points of the seven ICs are scaled up.
Regarding convergence, at a set-point scaling factor of 14, the proposed
formulation still converges, whereas the sequential method fails to converge.
Regarding computational time, the proposed formulation requires approximately one-eighth of the time of the sequential method, where each computational time is the average of 150 runs.
Compared with the sequential method, the proposed formulation therefore has
advantages in terms of both convergence and computational time.

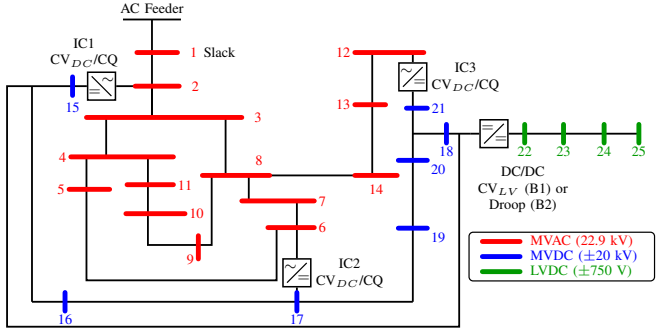
\begin{figure}[!t]
\centering
\resizebox{\columnwidth}{!}{%
\begin{tikzpicture}[
    font=\rmfamily\scriptsize,
    ac/.style={black, line width=0.9pt},
    dc/.style={black, line width=0.9pt},
    lv/.style={black, line width=0.9pt},
    busac/.style={red, line width=2.4pt, line cap=round},
    busdc/.style={blue, line width=2.4pt, line cap=round},
    buslv/.style={green!60!black, line width=2.4pt, line cap=round},
    dev/.style={draw=black, line width=0.75pt, fill=white,
        minimum width=0.48cm, minimum height=0.48cm, inner sep=0pt},
    labeltxt/.style={black, fill=white, inner sep=1.0pt},
    lblac/.style={red, fill=white, inner sep=1.0pt},
    lbldc/.style={blue, fill=white, inner sep=1.0pt},
    lbllv/.style={green!60!black, fill=white, inner sep=1.0pt},
    every node/.style={align=center}
]

\pgfdeclarelayer{busbars}
\pgfsetlayers{main,busbars}

\path[use as bounding box] (-0.20,-0.95) rectangle (11.35,5.05);

\newcommand{\vbusdc}[3]{%
    \begin{pgfonlayer}{busbars}\draw[busdc] (#1,#2-0.16) -- (#1,#2+0.16);\end{pgfonlayer}
    \node[lbllv, below=5pt] at (#1,#2) {#3};
}
\newcommand{\hbusdc}[3]{%
    \begin{pgfonlayer}{busbars}\draw[busdc] (#1-0.16,#2) -- (#1+0.16,#2);\end{pgfonlayer}
    \node[labeltxt, right=5pt] at (#1,#2) {#3};
}
\newcommand{\vbuslv}[3]{%
    \begin{pgfonlayer}{busbars}\draw[buslv] (#1,#2-0.16) -- (#1,#2+0.16);\end{pgfonlayer}
    \node[lbllv, below=5pt] at (#1,#2) {#3};
}
\newcommand{\vscicon}[3]{%
    \node[dev] (#1) at (#2,#3) {};
    \draw[black, line width=0.6pt] (#2-0.17,#3-0.17) -- (#2+0.17,#3+0.17);
    \node[black] at (#2-0.11,#3+0.10) {$\sim$};
    \node[black] at (#2+0.12,#3-0.10) {$=$};
}
\newcommand{\vsciconrev}[3]{%
    \node[dev] (#1) at (#2,#3) {};
    \draw[black, line width=0.6pt] (#2+0.17,#3-0.17) -- (#2-0.17,#3+0.17);
    \node[black] at (#2+0.11,#3+0.10) {$\sim$};
    \node[black] at (#2-0.12,#3-0.10) {$=$};
}
\newcommand{\dcdcicon}[3]{%
    \node[dev] (#1) at (#2,#3) {};
    \draw[black, line width=0.6pt] (#2-0.17,#3-0.17) -- (#2+0.17,#3+0.17);
    \node[black] at (#2-0.11,#3+0.10) {$=$};
    \node[black] at (#2+0.12,#3-0.10) {$=$};
}

\draw[black, line width=0.9pt] (2.05,4.55) -- (3.05,4.55);
\node[labeltxt, above=2pt] at (2.55,4.55) {AC Feeder};
\draw[ac] (2.55,4.55) -- (2.55,3.98);
\node[labeltxt, anchor=west] at (3.40,3.98) {Slack};

\draw[ac] (2.55,3.98) -- (2.55,2.86);
\begin{pgfonlayer}{busbars}\draw[busac] (2.29,3.98) -- (2.99,3.98);\end{pgfonlayer}
\node[lblac, right=5pt] at (2.99,3.98) {1};

\begin{pgfonlayer}{busbars}\draw[busac] (2.26,3.40) -- (3.02,3.40);\end{pgfonlayer}
\node[lblac, right=5pt] at (3.02,3.40) {2};

\begin{pgfonlayer}{busbars}\draw[busac] (1.42,2.86) -- (4.10,2.86);\end{pgfonlayer}
\node[lblac, right=5pt] at (4.10,2.86) {3};
\draw[ac] (2.55,3.40) -- (2.55,2.86);

\vsciconrev{ic1}{1.68}{3.40}
\node[labeltxt, above=1pt, xshift=-3mm] at (ic1.north) {IC1\\CV$_{DC}$/CQ};
\draw[ac] (ic1.east) -- (2.26,3.40);

\draw[ac] (1.76,2.86) -- (1.76,2.18);
\begin{pgfonlayer}{busbars}\draw[busac] (1.15,2.18) -- (2.92,2.18);\end{pgfonlayer}
\node[lblac, left=1pt] at (1.15,2.18) {4};

\draw[ac] (1.42,2.18) -- (1.42,1.62);
\begin{pgfonlayer}{busbars}\draw[busac] (1.12,1.62) -- (1.82,1.62);\end{pgfonlayer}
\node[lblac, left=1pt] at (1.12,1.62) {5};
\draw[ac] (1.42,1.62) -- (1.42,0.05) -- (4.70,0.05) -- (4.70,0.96);

\draw[ac] (2.48,2.18) -- (2.48,1.70);
\begin{pgfonlayer}{busbars}\draw[busac] (2.12,1.70) -- (2.92,1.70);\end{pgfonlayer}
\node[lblac, right=1pt] at (2.92,1.70) {11};

\draw[ac] (2.48,1.70) -- (2.48,1.20);
\begin{pgfonlayer}{busbars}\draw[busac] (2.10,1.20) -- (3.12,1.20);\end{pgfonlayer}
\node[lblac, right=1pt] at (3.12,1.20) {10};

\draw[ac] (2.48,1.20) -- (2.48,0.66) -- (3.35,0.66);
\begin{pgfonlayer}{busbars}\draw[busac] (3.35,0.42) -- (3.35,0.82);\end{pgfonlayer}
\node[lblac, left=1pt] at (3.35,0.42) {9};

\draw[ac] (3.82,2.86) -- (3.82,1.86);
\begin{pgfonlayer}{busbars}\draw[busac] (3.42,1.86) -- (4.58,1.86);\end{pgfonlayer}
\node[lblac, above=4pt] at (4.40,1.86) {8};
\draw[ac] (3.58,1.86) -- (3.58,0.66) -- (3.35,0.66);

\draw[ac] (4.22,1.86) -- (4.22,1.42);
\begin{pgfonlayer}{busbars}\draw[busac] (4.10,1.42) -- (5.35,1.42);\end{pgfonlayer}
\node[lblac, right=1pt] at (5.35,1.42) {7};
\draw[ac] (5.05,1.42) -- (5.05,0.96);
\begin{pgfonlayer}{busbars}\draw[busac] (4.55,0.96) -- (5.35,0.96);\end{pgfonlayer}
\node[lblac, right=1pt] at (5.35,0.96) {6};

\draw[ac] (4.58,1.86) -- (6.35,1.86);

\vscicon{ic2}{5.05}{0.18}
\node[labeltxt, right=1pt] at (ic2.east) (lbl2) {IC2\\CV$_{DC}$/CQ};
\draw[ac] (5.05,0.96) -- (ic2.north);

\draw[ac] (6.35,3.98) -- (6.35,1.86);

\begin{pgfonlayer}{busbars}\draw[busac] (6.05,3.98) -- (7.24,3.98);\end{pgfonlayer}
\node[lblac, left=1pt] at (6.05,3.98) {12};

\begin{pgfonlayer}{busbars}\draw[busac] (6.05,3.08) -- (6.58,3.08);\end{pgfonlayer}
\node[lblac, left=1pt] at (6.05,3.08) {13};

\begin{pgfonlayer}{busbars}\draw[busac] (6.05,1.86) -- (6.78,1.86);\end{pgfonlayer}
\node[lblac, below=3pt] at (6.42,1.86) {14};

\vscicon{ic3}{7.05}{3.56}
\node[labeltxt, right=1pt] at (ic3.east) (lbl3) {IC3\\CV$_{DC}$/CQ};
\draw[ac] (7.05,3.98) -- (ic3.north);

\draw[dc] (0.48,3.40) -- (0.48,-0.32);
\draw[dc] (0.48,3.40) -- (ic1.west);
\begin{pgfonlayer}{busbars}\draw[busdc] (1.18,3.24) -- (1.18,3.56);\end{pgfonlayer}
\node[lbldc, below=2pt] at (1.18,3.24) {15};

\draw[dc] (0.48,-0.32) -- (7.05,-0.32);
\begin{pgfonlayer}{busbars}\draw[busdc] (1.05,-0.48) -- (1.05,-0.16);\end{pgfonlayer}
\node[lbldc, below=5pt] at (1.05,-0.32) {16};

\draw[dc] (ic2.south) -- (5.05,-0.32);
\begin{pgfonlayer}{busbars}\draw[busdc] (5.05,-0.48) -- (5.05,-0.16);\end{pgfonlayer}
\node[lbldc, below=5pt] at (5.05,-0.32) {17};

\draw[dc] (7.05,-0.32) -- (7.05,3.02);
\draw[dc] (ic3.south) -- (7.05,3.02);

\begin{pgfonlayer}{busbars}\draw[busdc] (6.94,3.02) -- (7.30,3.02);\end{pgfonlayer}
\node[lbldc, right=1pt] at (7.30,3.02) {21};

\draw[dc] (0.48,3.40) -- (0.05,3.40) -- (0.05,-0.75) -- (7.85,-0.75) -- (7.85,2.58);

\draw[dc] (7.05,2.58) -- (8.05,2.58);
\begin{pgfonlayer}{busbars}\draw[busdc] (7.63,2.42) -- (7.63,2.74);\end{pgfonlayer}
\node[lbldc, below=5pt] at (7.63,2.58) {18};

\begin{pgfonlayer}{busbars}\draw[busdc] (6.80,2.12) -- (7.30,2.12);\end{pgfonlayer}
\begin{pgfonlayer}{busbars}\node[lbldc, below right=1pt] at (7.30,2.12) {20};\end{pgfonlayer}

\begin{pgfonlayer}{busbars}\draw[busdc] (6.80,0.95) -- (7.30,0.95);\end{pgfonlayer}
\node[lbldc, below right=1pt] at (7.30,0.95) {19};

\dcdcicon{dcdc}{8.45}{2.58}
\node[labeltxt, below=3pt] at (8.95,2.16) {DC/DC\\CV$_{LV}$ (B1) or\\Droop (B2)};
\draw[dc] (8.05,2.58) -- (dcdc.west);

\draw[lv] (dcdc.east) -- (10.95,2.58);
\vbuslv{8.98}{2.58}{22}
\vbuslv{9.65}{2.58}{23}
\vbuslv{10.35}{2.58}{24}
\vbuslv{10.95}{2.58}{25}

\draw[black, line width=0.6pt, rounded corners=2pt] (8.02,0.04) rectangle (11.22,0.88);
\begin{pgfonlayer}{busbars}\draw[busac] (8.25,0.72) -- (8.90,0.72);\end{pgfonlayer}
\node[labeltxt, text=red, right=4pt] at (8.90,0.72) {MVAC (22.9 kV)};
\begin{pgfonlayer}{busbars}\draw[busdc] (8.25,0.46) -- (8.90,0.46);\end{pgfonlayer}
\node[labeltxt, text=blue, right=4pt] at (8.90,0.46) {MVDC ($\pm$20 kV)};
\begin{pgfonlayer}{busbars}\draw[buslv] (8.25,0.20) -- (8.90,0.20);\end{pgfonlayer}
\node[labeltxt, text=green!60!black, right=4pt] at (8.90,0.20) {LVDC ($\pm$750 V)};

\end{tikzpicture}%
}
\caption{Modified CIGRE AC/DC distribution system.}
\label{fig:dis_grid}
\end{figure}
\begin{figure}[!t]
\centering
\definecolor{vMatC}{rgb}{0.02,0.08,0.38}
\definecolor{vTotC}{rgb}{0.38,0.70,1.00}
\definecolor{aMatC}{rgb}{0.45,0.00,0.00}
\definecolor{aTotC}{rgb}{1.00,0.55,0.50}
\definecolor{axV}{rgb}{0.12,0.42,0.80}
\definecolor{axA}{rgb}{0.75,0.15,0.15}
\resizebox{\columnwidth}{!}{%
\begin{tikzpicture}
\pgfplotsset{
  vMat/.style={vMatC, line width=1.0pt, mark=asterisk, mark size=1.6pt, mark options={solid, line width=0.6pt}},
  vTot/.style={vTotC, line width=1.0pt, dash pattern=on 3pt off 3pt, mark=asterisk, mark size=1.6pt, mark options={solid, line width=0.6pt}},
  aMat/.style={aMatC, line width=1.0pt, mark=triangle*, mark size=1.5pt, mark options={solid}},
  aTot/.style={aTotC, line width=1.0pt, dash pattern=on 3pt off 3pt, mark=triangle*, mark size=1.5pt, mark options={solid}},
}
\pgfplotsset{every axis/.append style={
    width=8.4cm, height=1.7cm, scale only axis,
    xmin=0.3, xmax=25.7, xtick={1,2,3,4,5,6,7,8,9,10,11,12,13,14,15,16,17,18,19,20,21,22,23,24,25},
    tick label style={font=\normalsize}, label style={font=\normalsize},
    title style={font=\normalsize, yshift=-3pt}}}

\begin{axis}[name=aV, at={(0,0)}, anchor=north west,
  axis y line*=left, ylabel={$|V|$ [p.u.]},
  ylabel style={axV}, yticklabel style={axV}, y axis line style={axV}, ytick style={axV},
  ymin=0.895, ymax=1.055, ytick={0.91,0.94,0.97,1.00,1.03},
  xticklabels={},
  xmajorgrids, ymajorgrids, grid style={gray!16, line width=0.3pt},
  title={(a) Scenario B1}]
\addplot[vMat] coordinates {(1,1.050311)(2,1.020894)(3,0.977978)(4,0.975116)(5,0.973531)(6,0.970147)(7,0.970361)(8,0.971973)(9,0.972261)(10,0.973594)(11,0.974198)(12,0.931377)(13,0.950656)(14,0.962652)(15,1.000069)(16,0.999329)(17,0.999970)(18,0.999830)(19,0.994293)(20,0.995639)(21,0.999949)(22,1.000000)(23,0.952527)(24,0.951578)(25,0.927819)};
\addplot[vTot] coordinates {(1,1.050000)(2,1.020676)(3,0.977642)(4,0.974770)(5,0.973182)(6,0.969787)(7,0.969994)(8,0.971623)(9,0.971915)(10,0.973256)(11,0.973864)(12,0.930927)(13,0.950278)(14,0.962319)(15,1.000000)(16,0.999304)(17,1.000000)(18,0.999874)(19,0.994333)(20,0.995684)(21,1.000000)(22,1.000000)(23,0.952529)(24,0.951580)(25,0.927822)};
\draw[densely dotted, black, line width=0.7pt] (axis cs:14.5,0.895)--(axis cs:14.5,1.055);
\draw[densely dotted, black!45, line width=0.6pt] (axis cs:21.5,0.895)--(axis cs:21.5,1.055);
\node[font=\small, anchor=north] at (axis cs:7.5,1.05) {MVAC};
\node[font=\small, anchor=north] at (axis cs:18,1.05) {MVDC};
\node[font=\small, anchor=north] at (axis cs:23.5,1.05) {LVDC};
\legend{}
\end{axis}
\begin{axis}[at={(0,0)}, anchor=north west, axis y line*=right, axis x line=none,
  ylabel={$\delta$ [deg]},
  ylabel style={axA}, yticklabel style={axA}, y axis line style={axA}, ytick style={axA},
  ymin=-6.3, ymax=0.65, ytick={-6,-4,-2,0}]
\addplot[aMat] coordinates {(1,0.000000)(2,-0.898512)(3,-2.295240)(4,-2.397934)(5,-2.460092)(6,-2.590552)(7,-2.579856)(8,-2.505041)(9,-2.488024)(10,-2.444551)(11,-2.425810)(12,-5.890844)(13,-4.216251)(14,-3.204839)};
\addplot[aTot] coordinates {(1,0.000000)(2,-0.894949)(3,-2.296778)(4,-2.399844)(5,-2.462248)(6,-2.593485)(7,-2.582765)(8,-2.507427)(9,-2.490349)(10,-2.446738)(11,-2.427939)(12,-5.905649)(13,-4.224628)(14,-3.209598)};
\end{axis}

\begin{axis}[name=bV, at={(0,-2.5cm)}, anchor=north west,
  axis y line*=left, ylabel={$|V|$ [p.u.]},
  ylabel style={axV}, yticklabel style={axV}, y axis line style={axV}, ytick style={axV},
  ymin=0.895, ymax=1.055, ytick={0.91,0.94,0.97,1.00,1.03},
  xticklabels={1,,3,,5,,7,,9,,11,,13,,15,,17,,19,,21,,23,,25},
  xticklabel style={font=\small},
  xmajorgrids, ymajorgrids, grid style={gray!16, line width=0.3pt},
  title={(b) Scenario B2},
  legend style={at={(0.5,-0.26)}, anchor=north, legend columns=4, font=\small,
                draw=black, fill=white, inner sep=1.5pt, /tikz/every even column/.append style={column sep=5pt}}]
\addlegendimage{vMat}\addlegendentry{EMT $|V|$}
\addlegendimage{vTot}\addlegendentry{Proposed $|V|$}
\addlegendimage{aMat}\addlegendentry{EMT $\delta$}
\addlegendimage{aTot}\addlegendentry{Proposed $\delta$}
\addplot[vMat] coordinates {(1,1.050311)(2,1.020890)(3,0.977968)(4,0.975106)(5,0.973521)(6,0.970137)(7,0.970351)(8,0.971962)(9,0.972250)(10,0.973583)(11,0.974187)(12,0.931353)(13,0.950639)(14,0.962639)(15,1.000070)(16,0.999329)(17,0.999970)(18,0.999830)(19,0.994292)(20,0.995638)(21,0.999949)(22,0.980356)(23,0.931768)(24,0.930751)(25,0.906431)};
\addplot[vTot] coordinates {(1,1.050000)(2,1.020670)(3,0.977629)(4,0.974756)(5,0.973168)(6,0.969772)(7,0.969980)(8,0.971608)(9,0.971900)(10,0.973242)(11,0.973849)(12,0.930896)(13,0.950255)(14,0.962300)(15,1.000000)(16,0.999304)(17,1.000000)(18,0.999873)(19,0.994333)(20,0.995684)(21,1.000000)(22,0.980356)(23,0.931768)(24,0.930752)(25,0.906432)};
\draw[densely dotted, black, line width=0.7pt] (axis cs:14.5,0.895)--(axis cs:14.5,1.055);
\draw[densely dotted, black!45, line width=0.6pt] (axis cs:21.5,0.895)--(axis cs:21.5,1.055);
\end{axis}
\begin{axis}[at={(0,-2.5cm)}, anchor=north west, axis y line*=right, axis x line=none,
  ylabel={$\delta$ [deg]},
  ylabel style={axA}, yticklabel style={axA}, y axis line style={axA}, ytick style={axA},
  ymin=-6.3, ymax=0.65, ytick={-6,-4,-2,0}]
\addplot[aMat] coordinates {(1,0.000000)(2,-0.898636)(3,-2.295525)(4,-2.398228)(5,-2.460389)(6,-2.590855)(7,-2.580160)(8,-2.505358)(9,-2.488336)(10,-2.444855)(11,-2.426110)(12,-5.892186)(13,-4.217063)(14,-3.205349)};
\addplot[aTot] coordinates {(1,0.000000)(2,-0.895094)(3,-2.297169)(4,-2.400251)(5,-2.462661)(6,-2.593912)(7,-2.583194)(8,-2.507865)(9,-2.490782)(10,-2.447159)(11,-2.428354)(12,-5.907437)(13,-4.225718)(14,-3.210291)};
\end{axis}

\end{tikzpicture}%
}
\caption{Comparison of the bus voltage magnitudes and AC-side voltage angles with a PSCAD EMT simulation: (a) Scenario B1 and (b) Scenario B2.}
\label{fig:cigre_validation}
\end{figure}
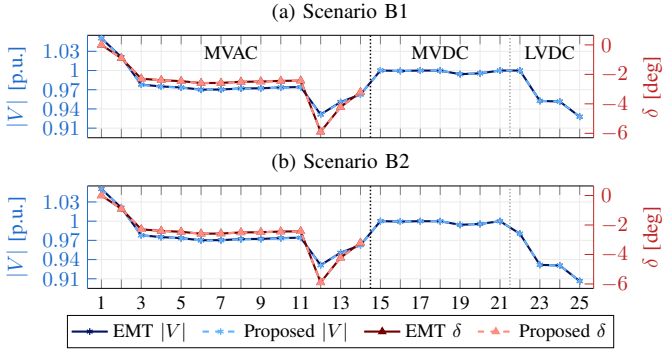

\subsection{Modified CIGRE AC/DC Distribution System}
\label{subsec:cigre}
For validation on a distribution system, the proposed formulation is applied
to a modified CIGRE AC/DC system~\cite{dognini2023service}.
Fig.~\ref{fig:dis_grid} shows the modified CIGRE AC/DC system, in which a
single slack bus at the point of connection to the transmission system feeds
a 22.9-kV MVAC network, three ICs link it to a $\pm$20-kV MVDC network,
and a DC/DC converter supplies a
$\pm$750-V LVDC feeder. Two scenarios (B1 and B2) are examined to test the generalized DC/DC operating-mode formulation, with all three ICs held in
$\mathrm{CV}_{DC}$/CQ mode. In Scenario B1, the DC/DC converter operates in the $\mathrm{CV}_{LV}$ mode, which fixes the LVDC voltage. Scenario B2 applies droop
control, in which the DC/DC converter varies its power injection with the
LVDC voltage. Both
scenarios are compared with a PSCAD electromagnetic-transient (EMT)
simulation of this system. The comparison shows the bus voltage
magnitudes across the three voltage levels, together with the AC-side
voltage angles. Fig.~\ref{fig:cigre_validation} shows that the solutions of the proposed
formulation are almost identical to the EMT reference, including the distinct LVDC voltage
profiles under the $\mathrm{CV}_{LV}$ and droop modes. The DC/DC converter model is thus validated to within $4.6 \times 10^{-4}$\,p.u. in magnitude and
$1.6 \times 10^{-2}\,\mathrm{deg}$ in angle.

\input{fig/fig_mg71_system}
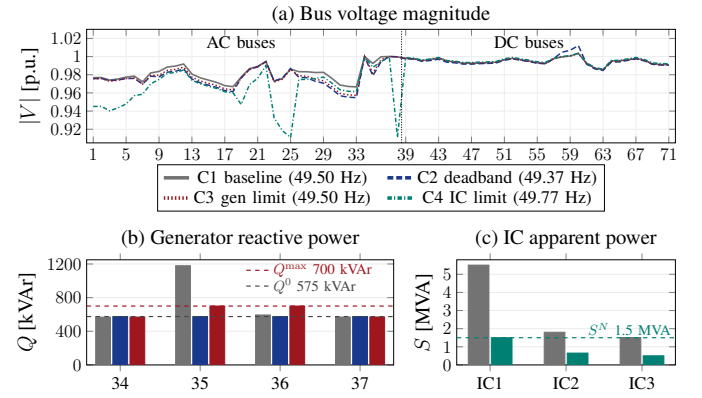
\begin{figure}[!t]
\centering
\definecolor{s1C}{rgb}{0.45,0.45,0.45}
\definecolor{s2C}{rgb}{0.10,0.22,0.55}
\definecolor{s3C}{rgb}{0.62,0.09,0.12}
\definecolor{s4C}{rgb}{0.00,0.50,0.45}
\resizebox{\columnwidth}{!}{%
\begin{tikzpicture}

\begin{axis}[name=volt, at={(0,0)}, anchor=north west,
  width=14.0cm, height=2.7cm, scale only axis,
  xmin=0.2, xmax=71.8,
  xtick={1,5,9,13,17,21,25,29,33,39,43,47,51,55,59,63,67,71},
  ymin=0.905, ymax=1.030, ytick={0.92,0.94,0.96,0.98,1.00,1.02},
  ylabel={$|V|$ [p.u.]}, label style={font=\Large}, tick label style={font=\large},
  xmajorgrids, ymajorgrids, grid style={gray!14, line width=0.3pt},
  legend style={at={(0.5,-0.19)}, anchor=north, legend columns=2, font=\large, draw=black,
                /tikz/every even column/.append style={column sep=8pt}},
                legend image post style={line width=2.4pt}]
\addplot[s1C, line width=1.1pt, mark=none] coordinates {(1,0.97684)(2,0.97691)(3,0.97414)(4,0.97534)(5,0.97742)(6,0.97849)(7,0.97221)(8,0.98204)(9,0.98363)(10,0.98848)(11,0.98955)(12,0.99185)(13,0.98078)(14,0.97669)(15,0.97413)(16,0.97164)(17,0.96799)(18,0.96688)(19,0.97754)(20,0.98670)(21,0.98926)(22,0.99450)(23,0.97303)(24,0.97586)(25,0.98622)(26,0.98338)(27,0.98328)(28,0.98246)(29,0.98265)(30,0.97624)(31,0.96875)(32,0.96711)(33,0.96661)(34,1.00008)(35,0.99245)(36,0.99976)(37,1.00007)(38,0.99973)(39,0.99800)(40,0.99758)(41,0.99556)(42,0.99683)(43,0.99858)(44,0.99500)(45,0.99426)(46,0.99280)(47,0.99264)(48,0.99305)(49,0.99324)(50,0.99377)(51,0.99664)(52,0.99800)(53,0.99650)(54,0.99502)(55,0.99320)(56,0.99258)(57,0.99766)(58,0.99971)(59,1.00061)(60,1.00350)(61,0.99252)(62,0.98724)(63,0.98553)(64,0.99502)(65,0.99522)(66,0.99654)(67,0.99800)(68,0.99589)(69,0.99203)(70,0.99124)(71,0.99105)};
\addlegendentry{C1 baseline (49.50 Hz)}
\addplot[s2C, line width=1.0pt, dash pattern=on 4pt off 2pt, mark=none] coordinates {(1,0.97574)(2,0.97582)(3,0.97348)(4,0.97422)(5,0.97579)(6,0.97595)(7,0.96972)(8,0.97843)(9,0.97776)(10,0.98281)(11,0.98393)(12,0.98630)(13,0.97527)(14,0.97119)(15,0.96864)(16,0.96616)(17,0.96251)(18,0.96140)(19,0.97651)(20,0.98579)(21,0.98862)(22,0.99452)(23,0.97299)(24,0.97636)(25,0.98684)(26,0.97719)(27,0.97665)(28,0.97278)(29,0.97073)(30,0.96438)(31,0.95693)(32,0.95530)(33,0.95480)(34,1.00005)(35,0.97884)(36,0.99429)(37,1.00005)(38,0.99976)(39,0.99746)(40,0.99725)(41,0.99477)(42,0.99570)(43,0.99709)(44,0.99382)(45,0.99310)(46,0.99173)(47,0.99167)(48,0.99221)(49,0.99243)(50,0.99299)(51,0.99603)(52,0.99746)(53,0.99596)(54,0.99448)(55,0.99266)(56,0.99204)(57,0.99786)(58,1.00475)(59,1.00696)(60,1.01212)(61,0.99173)(62,0.98644)(63,0.98473)(64,0.99390)(65,0.99417)(66,0.99578)(67,0.99746)(68,0.99535)(69,0.99149)(70,0.99069)(71,0.99050)};
\addlegendentry{C2 deadband (49.37 Hz)}
\addplot[s3C, line width=1.0pt, dash pattern=on 1pt off 1.5pt, mark=none] coordinates {(1,0.97569)(2,0.97576)(3,0.97288)(4,0.97395)(5,0.97588)(6,0.97642)(7,0.96961)(8,0.97935)(9,0.97942)(10,0.98483)(11,0.98595)(12,0.98835)(13,0.97729)(14,0.97320)(15,0.97064)(16,0.96815)(17,0.96450)(18,0.96339)(19,0.97645)(20,0.98607)(21,0.98880)(22,0.99434)(23,0.97192)(24,0.97512)(25,0.98590)(26,0.97889)(27,0.97840)(28,0.97502)(29,0.97333)(30,0.96695)(31,0.95949)(32,0.95785)(33,0.95735)(34,1.00007)(35,0.98173)(36,0.99662)(37,1.00007)(38,0.99972)(39,0.99799)(40,0.99757)(41,0.99555)(42,0.99682)(43,0.99858)(44,0.99499)(45,0.99426)(46,0.99280)(47,0.99263)(48,0.99305)(49,0.99324)(50,0.99376)(51,0.99663)(52,0.99799)(53,0.99649)(54,0.99501)(55,0.99319)(56,0.99257)(57,0.99765)(58,0.99970)(59,1.00060)(60,1.00349)(61,0.99251)(62,0.98723)(63,0.98553)(64,0.99502)(65,0.99521)(66,0.99653)(67,0.99799)(68,0.99588)(69,0.99202)(70,0.99123)(71,0.99103)};
\addlegendentry{C3 gen limit (49.50 Hz)}
\addplot[s4C, line width=1.0pt, dash pattern=on 3pt off 1.5pt on 0.8pt off 1.5pt, mark=none] coordinates {(1,0.94511)(2,0.94518)(3,0.94026)(4,0.94358)(5,0.94778)(6,0.95702)(7,0.95854)(8,0.97007)(9,0.97484)(10,0.98144)(11,0.98263)(12,0.98516)(13,0.97406)(14,0.96996)(15,0.96739)(16,0.96489)(17,0.96122)(18,0.96011)(19,0.94732)(20,0.96906)(21,0.97610)(22,0.99008)(23,0.93244)(24,0.91855)(25,0.91097)(26,0.97476)(27,0.97491)(28,0.97590)(29,0.97743)(30,0.97101)(31,0.96351)(32,0.96187)(33,0.96137)(34,1.00002)(35,0.98819)(36,0.99432)(37,0.99998)(38,0.91097)(39,0.99880)(40,0.99834)(41,0.99621)(42,0.99740)(43,0.99907)(44,0.99569)(45,0.99496)(46,0.99356)(47,0.99346)(48,0.99395)(49,0.99416)(50,0.99470)(51,0.99768)(52,0.99908)(53,0.99758)(54,0.99610)(55,0.99428)(56,0.99366)(57,0.99839)(58,1.00014)(59,1.00096)(60,1.00371)(61,0.99318)(62,0.98790)(63,0.98620)(64,0.99574)(65,0.99599)(66,0.99749)(67,0.99908)(68,0.99697)(69,0.99311)(70,0.99232)(71,0.99213)};
\addlegendentry{C4 IC limit (49.77 Hz)}
\draw[black, densely dotted, line width=0.7pt] (axis cs:38.5,0.905)--(axis cs:38.5,1.030);
\node[font=\large, anchor=north] at (axis cs:19,1.029) {AC buses};
\node[font=\large, anchor=north] at (axis cs:54,1.029) {DC buses};
\end{axis}
\node[font=\Large, anchor=south] at (volt.north) {(a) Bus voltage magnitude};
\begin{axis}[name=genP, at={(0,-5.48cm)}, anchor=north west,
  width=7.3cm, height=2.5cm, scale only axis,
  ybar=1.6pt, bar width=10pt, enlarge x limits=0.14,
  ymin=0, ymax=1250, ytick={0,400,800,1200}, yticklabel style={/pgf/number format/1000 sep={}},
  ylabel={$Q$ [kVAr]}, label style={font=\Large}, tick label style={font=\large},
  symbolic x coords={34,35,36,37}, xtick=data,
  ymajorgrids, grid style={gray!16},
  title={(b) Generator reactive power}, title style={font=\Large, yshift=-2pt}]
\addplot[fill=s1C, draw=s1C] coordinates {(34,568.3)(35,1179.3)(36,593.9)(37,569.5)};
\addplot[fill=s2C, draw=s2C] coordinates {(34,575.0)(35,575.0)(36,575.0)(37,575.0)};
\addplot[fill=s3C, draw=s3C] coordinates {(34,569.1)(35,700.0)(36,700.0)(37,569.7)};
\draw[gray!55!black, dashed, line width=0.8pt] ({rel axis cs:0,0}|-{axis cs:34,575})--({rel axis cs:1,0}|-{axis cs:34,575});
\draw[s3C, dashed, line width=0.8pt] ({rel axis cs:0,0}|-{axis cs:34,700})--({rel axis cs:1,0}|-{axis cs:34,700});
\draw[s3C, dashed, line width=0.8pt] (rel axis cs:0.52,0.90)--(rel axis cs:0.60,0.90);
\node[font=\normalsize,s3C,anchor=west,inner sep=1.5pt] at (rel axis cs:0.60,0.90) {$Q^{\max}$ 700 kVAr};
\draw[gray!55!black, dashed, line width=0.8pt] (rel axis cs:0.52,0.75)--(rel axis cs:0.60,0.75);
\node[font=\normalsize,gray!55!black,anchor=west,inner sep=1.5pt] at (rel axis cs:0.60,0.75) {$Q^0$ 575 kVAr};
\end{axis}

\begin{axis}[name=icP, at={(8.8cm,-5.48cm)}, anchor=north west,
  width=5.2cm, height=2.5cm, scale only axis,
  ybar=1.6pt, bar width=14pt, enlarge x limits=0.22,
  ymin=0, ymax=5.8, ytick={0,1,2,3,4,5},
  ylabel={$S$ [MVA]}, label style={font=\Large}, tick label style={font=\large},
  symbolic x coords={IC1,IC2,IC3}, xtick=data,
  ymajorgrids, grid style={gray!16},
  title={(c) IC apparent power}, title style={font=\Large, yshift=-2pt}]
\addplot[fill=s1C, draw=s1C] coordinates {(IC1,5.506)(IC2,1.803)(IC3,1.510)};
\addplot[fill=s4C, draw=s4C] coordinates {(IC1,1.500)(IC2,0.653)(IC3,0.503)};
\draw[s4C, dashed, line width=0.8pt] ({rel axis cs:0,0}|-{axis cs:IC1,1.5})--({rel axis cs:1,0}|-{axis cs:IC1,1.5});
\node[font=\normalsize,s4C,anchor=south east,inner sep=1.5pt] at ({rel axis cs:0.99,0}|-{axis cs:IC1,1.5}) {$S^{N}$ 1.5 MVA};
\end{axis}

\end{tikzpicture}%
}
\caption{Scenarios C1--C4 on the islanded AC/DC microgrid: (a) bus voltage magnitude, (b) generator reactive power, and (c) IC apparent power.}
\label{fig:mg71_validation}
\end{figure}

\subsection{Islanded AC/DC Microgrid}
\label{subsec:mg71}
The formulation is applied to a modified islanded AC/DC
hybrid microgrid~\cite{surendra2024static} with no slack bus, shown in Fig.~\ref{fig:mg71_system}. In this system, a 38-bus
12.67-kV AC microgrid and a 33-bus 20.67-kV DC microgrid are interlinked by
three ICs. The generators operate as DGs in droop,
and the ICs operate in both active- and reactive-power droop (Fig.~\ref{fig:mg71_system} details the droop control settings for each component). The baseline Scenario C1 is compared with three scenarios, each obtained by applying one outer-loop element to C1, namely the generator droop deadband (C2), the generator reactive-power limit (C3), and the IC output limit (C4). In all four scenarios, each operating mode
undergoes at most one transition, without oscillation. Fig.~\ref{fig:mg71_validation}(a) shows the bus voltage magnitudes. The AC-side voltages drop as the droop deadband or an output limit restricts the
voltage support from the generators or ICs. Since the AC microgrid has no slack
bus, its frequency is solved as a state variable, and its value in each scenario is
given in the legend of Fig.~\ref{fig:mg71_validation}(a).
Fig.~\ref{fig:mg71_validation}(b) shows the AC-side generators' reactive-power
outputs, which the deadband holds at their set-points in C2, while two of the
constrained generators reach their limits in C3.
Fig.~\ref{fig:mg71_validation}(c) shows the apparent-power injections of the ICs, with
IC1 operating at its apparent-power capability in C4. The unified NR formulation therefore
enforces the IC output limit.

Table~\ref{tab:tools} compares the capabilities of existing
methods~\cite{zulkifli2026generalized,thurner2018pandapower,bolgaryn2024further,ergun2019optimal} and
the proposed formulation in handling the features examined in the case studies. Only the proposed formulation implements every feature across all scenarios.

\subsection{Additional Case Studies}
\label{subsec:additional}
Additional case studies are conducted for three purposes: (i) to validate the
three-winding AC transformer model using an AC/DC system adapted from
pandapower's multi-voltage network, (ii) to evaluate scalability using a modified
2869-bus \mbox{PEGASE} AC/DC system, and (iii) to show the convergence
and computational-time benefits of integrating outer-loop elements into the
unified formulation using the AC/DC microgrid in Section~\ref{subsec:mg71}.
Details are provided in Appendix C.

\section{Conclusion}
\label{sec:conclusion}
This paper has presented a generalized and unified NR power flow formulation for
AC/DC hybrid grids. The generalized models represent the AC and DC networks and
loads as well as the generators, ICs, and voltage-transformation devices. In
particular, the operating modes of generators and converters are expressed
through a common set of coefficients. For the unified NR formulation,
augmented variables and corresponding constraints are introduced to resolve
the outer-loop elements together with the network equations. Case studies validate the proposed formulation across AC/DC hybrid grids with diverse operating conditions and network structures, ranging from transmission and distribution networks to islanded microgrids.
The implementation is made publicly available to facilitate its use in analyzing diverse AC/DC hybrid
grids~\cite{chang2026unigrid}. The proposed formulation is thus expected to
contribute to the analysis of future power systems with mixed AC/DC networks
and increasing integration of power electronic devices.

\begin{table}[t]
\centering
\caption{Features Implemented in Unified Power Flow Methods}
\label{tab:tools}
\scriptsize
\setlength{\tabcolsep}{5.4pt}
\renewcommand{\arraystretch}{1.2}
\begin{tabular}{@{}lcccccccc@{}}
\hline
 & \multicolumn{2}{c}{\textbf{Section IV-A}} & \multicolumn{2}{c}{\textbf{Section IV-B}} & \multicolumn{4}{c}{\textbf{Section IV-C}} \\
\cline{2-3}\cline{4-5}\cline{6-9}
\multicolumn{1}{@{}r}{\textbf{Scenario}} & \textbf{A1} & \textbf{A2} & \textbf{B1} & \textbf{B2} & \textbf{C1} & \textbf{C2} & \textbf{C3} & \textbf{C4} \\
\hline
VeraGrid~\cite{zulkifli2026generalized} & $\bigcirc$ & $\times$ & $\times$ & $\times$ & $\times$ & $\times$ & $\bigcirc$ & $\bigcirc$ \\
pandapower~\cite{thurner2018pandapower,bolgaryn2024further} & $\bigcirc$ & $\times$ & $\times$ & $\times$ & $\times$ & $\bigcirc$ & $\bigcirc$ & $\times$ \\
PowerModelsACDC~\cite{ergun2019optimal} & $\bigcirc$ & $\bigcirc$ & $\times$ & $\times$ & $\times$ & $\times$ & $\times$ & $\bigcirc$ \\
\textbf{Proposed} & $\bigcirc$ & $\bigcirc$ & $\bigcirc$ & $\bigcirc$ & $\bigcirc$ & $\bigcirc$ & $\bigcirc$ & $\bigcirc$ \\
\hline
\end{tabular}
\end{table}

\enlargethispage{\baselineskip}
\section*{Appendix}
The Appendix of this paper is included at the end of this preprint and is also available in~\cite{chang2026appendix}.
Appendices A, B, and C present the three-winding transformer model, the Jacobian
derivation, and additional case studies, respectively.

\bibliographystyle{IEEEtran}
\bibliography{refs}

\clearpage
\includepdf[pages=-]{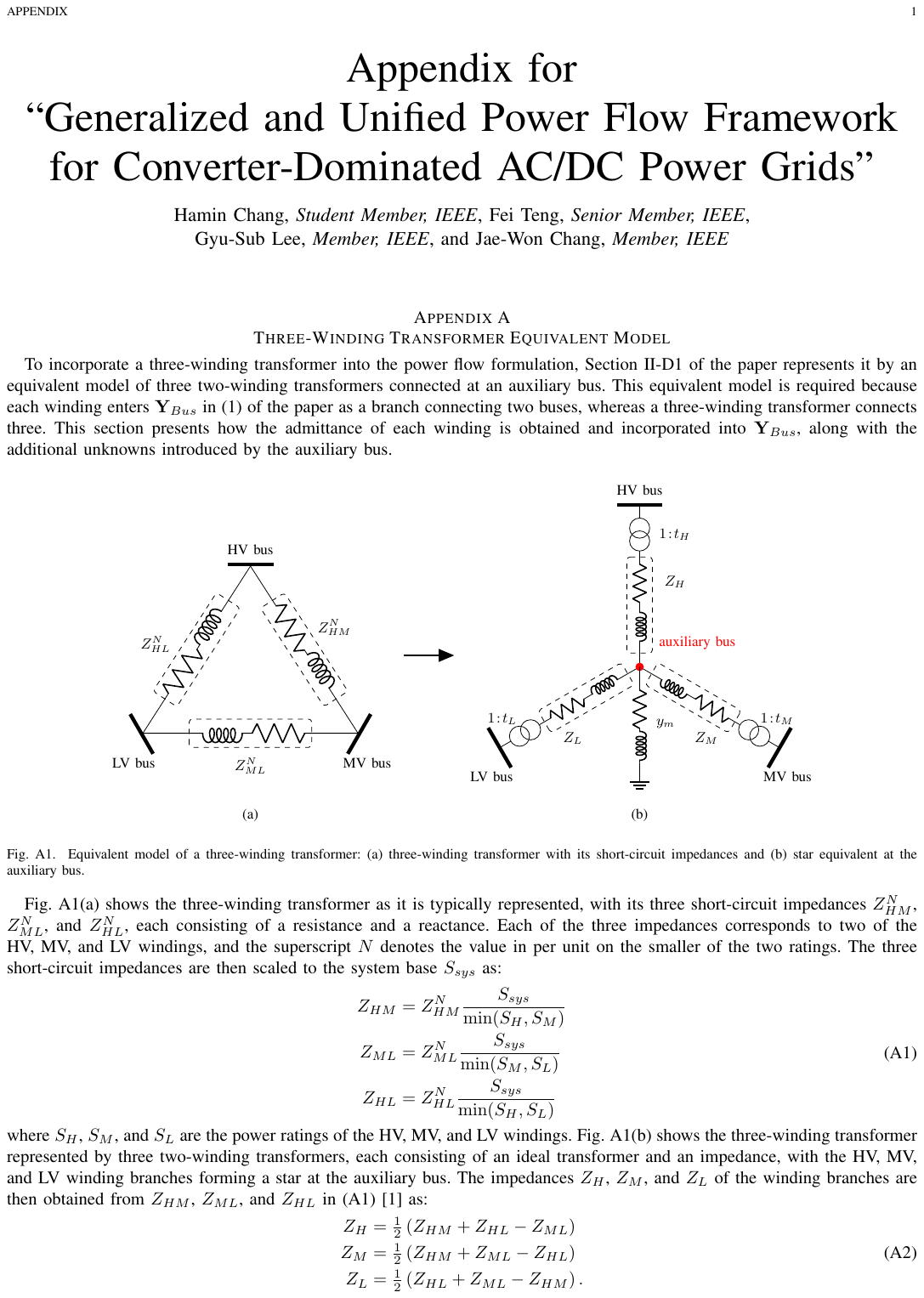}
\end{document}